\documentclass{aastex62}
\pdfoutput=1
\usepackage{graphicx}
\usepackage{subfigure}
\usepackage{multirow}
\usepackage{comment}
\usepackage{natbib}
\usepackage{hyperref}
\usepackage{mathtools}
\usepackage{mathrsfs}
\usepackage{fontenc}
\usepackage{color}
\usepackage{url}
\usepackage{hyperref}
\usepackage{gensymb}
\usepackage{pifont}
\newcommand       \Angstrom     {\,{\rm \AA}}

\newcommand       \yr       {\,{\rm yr}}

\newcommand       \mum          {\,{\rm \mu m}}

\newcommand{\spitzerirs}{{\em Spitzer}/IRS\ } 

\newcommand{\etal}{\textrm{et al.\ }}

\newcommand{\eg}{\textrm{e.g., }}

\begin{document}

\title{A New Calibration of Star Formation Rate in Galaxies Based on \\
         Polycyclic Aromatic Hydrocarbon Emission}

\shorttitle{SFR Traced by PAHs}

\author{Yanxia Xie}
\affil{Kavli Institute for Astronomy and Astrophysics, Peking University,
Beijing 100871, China {\sf yanxia.xie@pku.edu.cn}}

\author{Luis C. Ho}
\affil{Kavli Institute for Astronomy and Astrophysics, Peking University,
Beijing 100871, China {\sf lho.pku@gmail.com}}
\affil{Department of Astronomy, School of Physics, Peking University,
Beijing 100871, China}

%\ {\sf yanxia.xie@pku.edu.cn}  \ {\sf lho.pku@gmail.com} 

\begin{abstract} 
Polycyclic aromatic hydrocarbon (PAH) emission has long been proposed to be a potential star formation rate indicator, as it arises from the photodissociation region bordering the Str\"omgren sphere of young, massive stars.  We apply a recently developed technique of mid-infrared spectral decomposition to obtain a uniform set of PAH measurements from {\it Spitzer}\ low-resolution spectra of a large sample of star-forming galaxies spanning a wide range in stellar mass ($M_\star \approx 10^6-10^{11.4}\,M_{\odot}$) and star formation rate ($\sim 0.1- 2000\,M_{\odot}\rm\,yr^{-1}$).  High-resolution spectra are also analyzed to measure [Ne~II] 12.8 $\mu$m and [Ne~III] 15.6 $\mu$m, which effectively trace the Lyman continuum.  We present a new relation between PAH luminosity and star formation rate based on the [Ne~II] and [Ne~III] lines.  Calibrations are given for the integrated 5--15 $\mu$m PAH emission, the individual features at 6.2, 7.7, 8.6, and 11.3 $\mu$m, as well as several mid-infrared bandpasses sensitive to PAH.  We confirm that PAH emission is suppressed in low-mass, dwarf galaxies, and we discuss the possible physical origin of this effect.
\end{abstract} 

\keywords{galaxies: formation --- galaxies: starburst --- galaxies: dwarf --- (ISM:) dust, extinction}

\section{Introduction}\label{sec:intro} 

The star formation rate (SFR) is vital to understand how galaxies acquire their mass, but it is a challenging quantity to measure in external galaxies, wherein it can only be inferred through indirect estimators.  Dust-corrected ultraviolet (UV) continuum emission and optical line emission ([O~II] $\lambda$3727, $\mathrm H{\alpha}$) have been used frequently to probe star formation on timescales of a few to $\sim$200 Myr (\eg Gallagher et al. 1989; Kennicutt 1998; Hao et al. 2011; Kennicutt \& Evans 2012).  These UV and optical tracers, however, suffer sensitively from dust extinction and are unavailable for highly obscured systems, such as ultraluminous infrared (IR) galaxies (ULIRGs). Moreover, the interstellar environments of some systems (e.g., dwarf galaxies and outer disks of spirals) may experience severe Lyman continuum photon leakage, posing a major complication for SFR indicators based on recombination lines (Hunter et al. 2010; Rela{\~n}o \etal 2012; Calzetti 2013). 

Dust absorbs the ionizing continuum from massive stars and reradiates the energy in the IR.  Thus, barring situations where IR cirrus emission heated by old stars dominates (Kennicutt \etal 2009), or in situations where dust opacity drops toward young stars (Calzetti 2001), the total IR luminosity ($L_{\mathrm{IR}}$) effectively probes the SFR on timescales of $\mathrm \sim 10^{8}\, \yr$.  The full IR spectral energy distribution is not always accessible for most galaxies, and, in practice, the monochromatic luminosity in a single band is often adopted as a SFR indicator (e.g., 24$\mum$: Alonso-Herrero \etal 2006, P{\'e}rez-Gonz{\'a}lez \etal 2006, Rela{\~n}o \etal 2007, Rieke \etal 2009; 70 and 160$\mum$: Bavouzet \etal 2008, Calzetti \etal 2010).  Furthermore, the IR window contains several prominent fine-structure lines that have been proposed as SFR estimators.  As discussed by Ho \& Keto (2007; see also Zhuang et al. 2019), the fine-structure lines [Ne~II] 12.8$\mum$ and [Ne~III] 15.6$\mum$ are especially promising.  [C~II] 158$\mum$, a major coolant associated with photodissociation regions (PDRs), has long been recognized as an important tracer of recent star formation (\eg Stacey \etal 1991, 2010; De Looze \etal 2011; Vallini \etal 2015), its importance highlighted all the more with the advent of the Atacama Large Millimeter/sub-millimeter Array, which can detect [C~II] out to very high redshifts (e.g., Pentericci et al. 2016; Brada\v c et al. 2017). 

Similar to [C~II] 158$\mum$, the mid-IR emission features from polycyclic aromatic hydrocarbons (PAHs) also arise from PDRs and hence potentially can serve as a SFR indicator, as attested by the empirical correlation between PAH emission and other star formation tracers (\eg F\"orster Schreiber \etal 2004; Peeters \etal 2004).  The PAH bands are rich, pervasive, and energetically significant, dominating the mid-IR spectrum from $\sim$3 to 20$\mum$ and accounting for $\sim$10\% of the IR energy budget in starburst galaxies (\eg Smith \etal 2007; Shipley \etal 2013).  These attributes render PAH emission a powerful tool for probing star formation over a wide range of redshifts.  In view of these advantages, there have many previous efforts to devise a quantitative calibration of the SFR in star-forming galaxies based on the strength of PAH emission (\eg Wu \etal 2005; Farrah \etal 2007; Pope \etal 2008; Treyer \etal 2010; Diamond-Stanic \& Rieke 2012; Shipley \etal 2016). 

This study offers a new PAH-based SFR calibration, using a uniform set of PAH measurements derived from a technique recently introduced by Xie et al. (2018).  Our technique utilizes a theoretical PAH template calculated from Draine \& Li (2007) to decouple the complex PAH features from other dust components in the mid-IR spectral region covered by the \textit{Spitzer} Infrared Spectrograph (IRS; Houck \etal 2004; Werner et al. 2004), allowing us to obtain robust measurements of the integrated PAH emission, including the associated PAH continuum.  We present calibrations for the total (5--15 $\mu$m) PAH luminosity as well as for the luminosity of certain key individual PAH bands.  Our calibration is anchored by the relationship between SFR and the strength of the mid-IR lines of [Ne~II] 12.8$\mum$ and [Ne~III] 15.6$\mum$, as originally proposed by Ho \& Keto (2007) and recently revisited by Zhuang \etal (2019).  The common bandpass of PAH and neon emission minimizes complications due to dust obscuration and aperture mismatch.  We show that this approach produces SFRs that are generally consistent with those derived from the IR continuum, extinction-corrected $\rm H{\alpha}$, or from the combination of $\rm H\alpha + 24\mum$. Special care must be taken, however, to treat dwarf galaxies, which tend to show a deficit in PAH.
 
We introduce the data in Section~\ref{sec:sample_data}. The measurement of the PAH and neon lines are given in Section~\ref{sec:measure}. Section 4 presents the new SFR calibration, which is discussed in Section~\ref{sec:discussion}. Readers who are only interested in the final calibration can go directly to Section~\ref{sec:result}. A summary is given in Section~\ref{sec:summary}. Throughout this paper, we adopt the cosmological parameters $\Omega_m = 0.308$, $\Omega_\Lambda = 0.692$, and $H_{0}=67.8$ km s$^{-1}$ Mpc$^{-1}$  for a $\Lambda$CDM cosmology (Planck Collaboration et al. 2016).  For galaxies with $z<0.02$, the luminosity distance is calculated from the heliocentric velocity corrected for the flow-field model of Mould \etal (2000).

\section{Calibration Sample \label{sec:sample_data}}

\subsection{Galaxy Selection}

Our calibration data are gathered from the literature and from the {\it Spitzer}\ archive.  The sample comprises local dwarf galaxies, normal star-forming galaxies, and starburst galaxies that span a large dynamical range in luminosity, stellar mass, and SFR. We require the selected galaxies to have high-quality IRS low-resolution and high-resolution spectra, the former providing the broad spectral coverage necessary to cover the widespread PAH emission and the latter having sufficient resolution to properly deblend [Ne~II]~12.8$\mum$ and [Ne~III]~15.6$\mum$ from neighboring contaminating features. After extensive experimentation with the IRS database, we determined that reliable measurements can be obtained from spectra that have a minimum signal-to-noise ratio per pixel of $\sim 1$.

Our calibration sample consists of three subsets of massive star-forming galaxies, supplemented with a subsample of dwarf galaxies to extend the dynamic range in physical properties, as follows:

\begin{enumerate}
\item{The Spitzer-SDSS-GALEX Spectroscopic Survey (SSGSS) of the SWIRE/Lockman Hole region provides 65 local ($0.03<z<0.21$) star-forming galaxies spanning 2 orders of magnitude in stellar mass, color, and dust attenuation (O'Dowd \etal 2009, 2011). SSGSS contains short-low and long-low IRS spectra, but only a subset of the 21 brightest galaxies has high-quality short-high spectra.}

\item{The Spitzer SDSS Statistical Spectroscopic Survey (S5), an expanded version of SSGSS, offers 175 additional star-forming galaxies at $0.05<z<0.1$ with suitable low- and high-resolution spectra.  The S5 galaxies are selected to have prior photometric coverage in the UV and $\mathrm {H\alpha}$ fluxes greater than $3\times10^{-15}\, {\rm erg\,s^{-1}\,cm^{-2}}$. Combining SSGSS and S5 results in 196 normal star-forming galaxies, which cover total IR luminosities $L_{\rm IR} \approx 10^9-10^{11}\,L_{\odot}$ and stellar masses $M_\star \approx 10^9-10^{11}\,M_{\odot}$\footnote{The stellar masses are taken from the MPE-JHU value-added catalog for the 7th data release of SDSS({https://wwwmpa.mpa-garching.mpg.de/SDSS/DR7/\#derived}).}.}

\item{To incorporate more luminous, more massive objects, we add IR-luminous galaxies from Farrah \etal (2007), who presented 53 ULIRGs observed with IRS.  As we are concerned with objects powered by star formation, we only retain the 13 objects that show no obvious signs of strong active galactic nuclei based on diagnostics in the X-rays (Iwasawa \etal 2011; Torres-Alb{\`a} \etal 2018), optical (Veilleux \etal 1995, 1999), and IR (Armus \etal 2007; Farrah \etal 2007).  Among them, three have stellar mass measurements in Shangguan \etal (2018); for the remaining 10, we estimate stellar masses from the 2MASS $J$-band photometry (see  Appendix~\ref{appendix:mstar_bcd_ulirg}) following the methodology of Shangguan \etal (2018).  The ULIRGs span $L_{\rm IR} \approx 10^{12.0}-10^{12.9} \,L_{\odot}$ and $M_\star \approx 10^{10.4}-10^{11.4}\,M_{\odot}$.}

\item{At the opposite extreme, we include less massive, low-luminosity, low-metallicity dwarf galaxies.  A systematic search of the {\it Spitzer}/IRS archive uncovered 18 dwarf galaxies with suitable observations, 17 of which we retain because either PAH or neon is detected.  They are all considered blue compact dwarfs (BCDs), including two of the most metal-poor galaxies known---I~Zw~18 with metallicity $Z=1/50\, Z_\odot$ (Searle \& Sargent 1972) and SBS~0335$-$052 with $Z= 1/41\, Z_\odot$ (Izotov \etal 1997). The MPE-JHU catalog gives stellar masses for four of the objects, and for the remaining 14 we calculate stellar masses according to the prescription of Bell \etal (2003), using their $J$-band luminosity and $\langle g-i \rangle$ color (Appendix~\ref{appendix:mstar_bcd_ulirg}). The stellar masses of the BCDs range mostly between $M_\star \approx 10^6$ and $10^9\,M_{\odot}$, except for  Haro~11 and CG~0752, which are distinctly more massive ($M_\star \approx 10^{10}\,M_{\odot}$).  Haro~11 qualifies as a luminous IR galaxy (Wu \etal 2006), but it is reported to have very low metallicity ($Z=1/10 \, Z_\odot$; Bergvall \etal 2000).}

\end{enumerate}

\begin{figure}%[ht]
\begin{center}
\includegraphics[height=0.4\textheight]{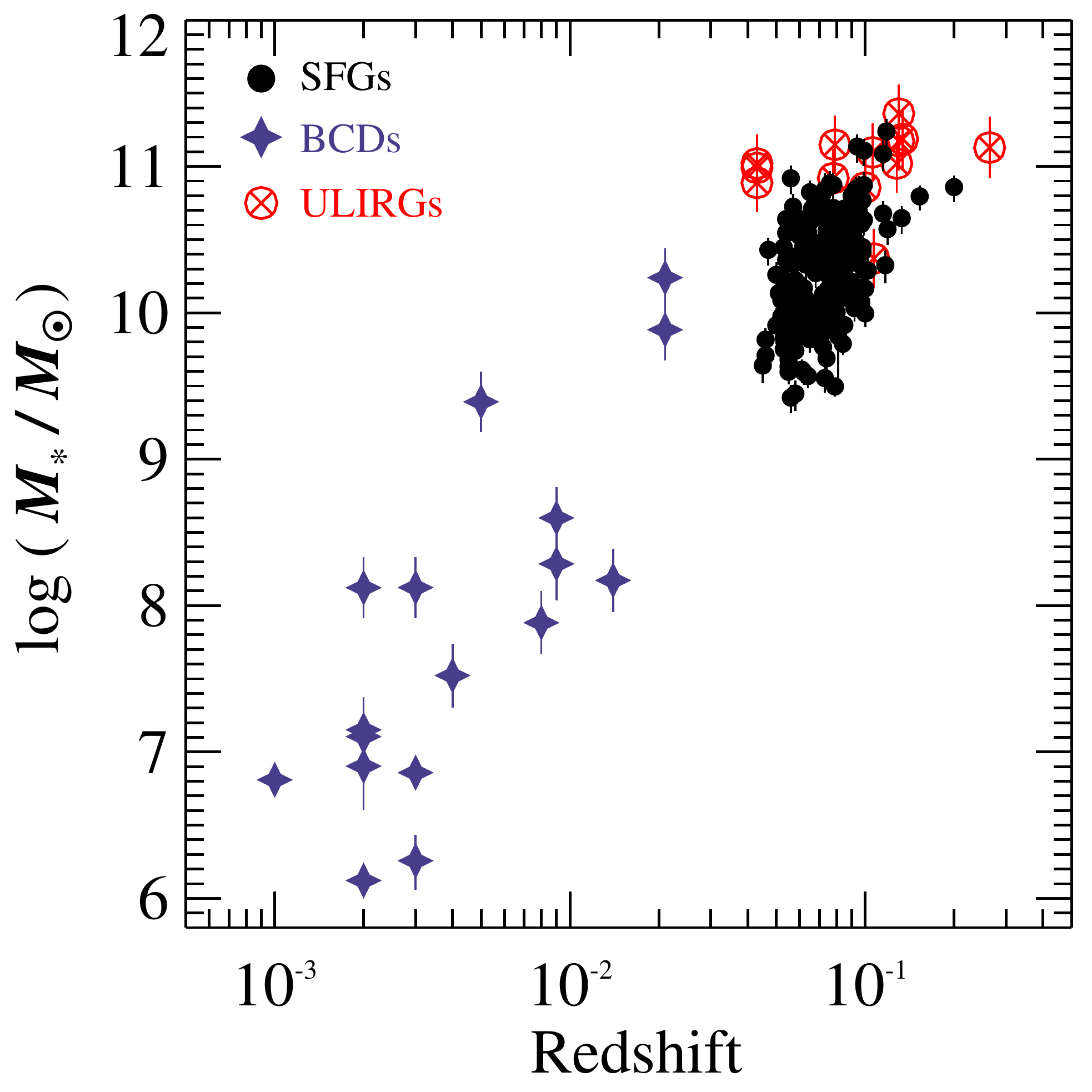}
\caption{Distribution of redshift and stellar mass of the sample, which comprises star-forming galaxies (SFGs), blue compact dwarfs (BCDs), and ultraluminous IR galaxies (ULIRGs).}
\label{fig:basic_par}
\end{center}
\end{figure}

Figure~\ref{fig:basic_par} summarizes the redshift and stellar mass distribution of the combined sample of 226 galaxies (196 normal star-forming galaxies, 13 ULIRGs, and 17 BCDs; Table~\ref{tab:main}). The entire sample covers $z<0.3$ and spans $M_\star \approx 10^6-10^{11.4}\,M_{\odot}$.

\subsection{Data}

The IRS has both a low-resolution and a high-resolution mode, each with short and long slits to cover a different wavelength range.  The short-low mode has a slit size of $3.7^{\prime\prime}\times 57^{\prime\prime}$ and $3.6^{\prime\prime} \times 57^{\prime\prime}$, covering, respectively, $7.4-14.5 \mum$ and $5.2-7.7 \mum$, while the long-low mode covers  $19.5-38.0 \mum$ with the $10.7^{\prime\prime}\times 168^{\prime\prime}$ slit and $14.0-21.3 \mum$ with the $10.5^{\prime\prime}\times 168^{\prime\prime}$ slit. The resolution varies from $\lambda/\Delta\lambda \approx 64$ to 128 in each segment. The short-high mode samples $9.9 - 19.6\mum$ with the $4.7^{\prime\prime} \times 11.3^{\prime\prime}$ slit, the long-high mode $18.7 - 37.2\mum$ with the $11.1^{\prime\prime} \times 22.3^{\prime\prime}$ slit, both with $\lambda/\Delta\lambda \approx 600$.  We give priority to spectra archived in the Cornell AtlaS of {\it Spitzer}/IRS Sources (CASSIS; Lebouteiller \etal 2011, 2015), which adopts a spectral extraction scheme according to the spatial extent of each source.  Sources not included in CASSIS were retrieved from the Spitzer Legacy Program Database\footnote{https://irsa.ipac.caltech.edu/data/SPITZER/S5/, https://irsa.ipac.caltech.edu/data/SPITZER/SSGSS/}.  

Aperture mismatch is a major concern given the relative proximity of our sources and the variety of slit widths employed by IRS.  We adopt the following strategy to mitigate this complication.  Recalling that our primary goal is to calibrate PAH emission relative to the independent SFR indicator based on [Ne~II]~12.8$\mum$ and [Ne~III]~15.6$\mum$, we note that the bulk of the PAH emission is confined to the $\sim 5-15 \mum$ region covered by the short-low spectrum (see Figure~\ref{fig:pah_filter}a), while both neon lines are contained in the short-high spectrum.  Fortunately, the apertures of both modes are quite similar, $3.7^{\prime\prime}$ for short-low and $4.7^{\prime\prime}$ for short-high, and for the purposes of this study, we will assume that the difference is unimportant.  For the normal star-forming galaxy (SSGSS and S5) sample, aperture correction has already been performed for the high- and low-resolution spectra (\eg O'Dowd et al. 2011).  As for the rest, the ULIRGs with $z \ga 0.08$ are sufficiently compact to be fully captured by the short-low aperture. Four ULIRGs with $z < 0.08$ and most of the BCDs (median $z = 0.0035$) are of possible concern, but in Section 4.1 we conclude that any residual aperture effects introduced by these objects do not affect our final conclusions.

The PAH decomposition method that we use (Xie et al. 2018) requires that the short-low spectrum be combined with the long-low spectrum to provide a sufficiently wide wavelength coverage extending to $\sim 38 \mum$.  We scale the larger aperture ($10.7^{\prime\prime}$) long-low spectrum to match the short-low spectrum using their common overlapping region, making the explicit assumption that the mid-IR spectrum does not change dramatically between these two apertures. This generally holds for nearby galaxies that have been studied with spatially resolved mid-IR observations (\eg Maragkoudakis \etal 2018). 

\section{Measurements \label{sec:measure}}

\subsection{Neon Lines \label{subsec:neon}}

The [Ne~II]\,12.8$\mum$ and [Ne~III]\,15.6$\mum$ lines, being narrow and well-separated from nearby features in the short-high spectrum, can be measured straightforwardly using a single Gaussian fit on top of a local continuum.  We select continuum regions immediately adjacent to either side of each line and interpolate with a linear continuum.  The distribution of repeated observations gives the flux and its uncertainty.  To quantify the significance of the line measurement, we define the quantity $S_{\mathrm {Ne}}$ as the ratio of the line flux to the full width at half maximum (FWHM) of the line.  We consider a line detected if $S_{\mathrm {Ne}} \geq 3 \sigma$, where $\sigma$ is the standard deviation of the local continuum; otherwise, the line is a non-detection, and the $3 \sigma$ upper limit is given by fixing the line width to the median value of the detected sources (FWHM = 1375 km~s$^{-1}$ for [Ne~II] and 1133 km~s$^{-1}$ for [Ne~III]).
In total, there are 7 upper limits for [Ne~II] and 100 upper limits for [Ne~III]. The total flux of neon is the sum of [Ne~II] and [Ne~III], and neon is regarded as detected when either line is detected.  Errors are estimated from boostrap sampling.
Two sources are undetected in neither [Ne~II] nor [Ne~III].  The extinction-corrected neon luminosities are given in Table~\ref{tab:main}.

\subsection{PAH Features \label{subsec:pah}} 

We measure the PAH strength using the methodology of Xie \etal (2018), which is summarized briefly here.   After first removing ionic emission lines that are not blended with the main PAH features, we fit the $\sim 5-38 \mum$ spectrum with a four-component model consisting of a theoretical PAH template plus dust continuum represented by three modified blackbodies of different temperatures, all subject to dust attenuation by foreground extinction. Our theoretical PAH spectrum is calculated by adopting a starlight intensity $U$\,=\,1 times the interstellar radiation field of the solar neighborhood interstellar medium (Mathis \etal 1983) and grain sizes $3.5\Angstrom < a < 20\Angstrom$ to account for stochastic heating of small grains (Draine \& Li 2001). The PAH spectrum is relatively insensitive to radiation intensities up to $U \approx 10^4$.  The global best fit is determined using the Levenberg-Marquardt  $\chi^2$-minimization algorithm {\tt MPFIT} (Markwardt 2009).  To estimate the uncertainties of the final PAH spectra, we randomly sample the observed spectra 100 times and repeat the fits; the median and standard deviation of the 100 realizations give the final PAH flux and its $1 \sigma$ error.  Upper limits (which, in this sample, apply only to some of the BCDs) on the PAH flux are set as $3 \sigma$.  We further consider the systematic uncertainties associated with the continuum model, which are estimated from Monte Carlo simulations as described in Xie \etal (2018; their Section~2 and Figure~2).  The final error budget on the PAH flux is the quadrature sum of these two contributions.

Table~\ref{tab:lumpah} lists the integrated $5-15 \mum$ PAH flux, as well as the flux of the commonly used individual PAH bands at 6.2, 7.7, 8.6 and 11.3$\mum$, adopting a Drude profile and the parameters given in Table~1 of Draine \& Li (2007).  To obtain uncertainties for each feature, we repeat the fitting 100 times on randomly simulated PAH spectra and take the median and standard deviation as the final flux and associated error. Dust extinction can be significant, especially for ULIRGs, where silicate absorption at 9.7$\mum$ can affect the 8.6 and 11.3$\mum$ PAH features by more than a factor of 10.  Mid-IR extinction is generally negligible ($< 2\%$) for ordinary star-forming galaxies.
 
%\input{table_basic_long.tex}
%%%%%%%%%%%%%%%%%%%%%%%%%%%%%%%%%%%%%%%%%%%%%%%
%Edited by LCH, April 17, 2019
%Edited by LCH, Aug 1, 2019
%
\begin{longrotatetable}
\begin{deluxetable}{l c c c c c c c c c c c}
\tablecaption{Physical Properties of the Sample \label{tab:main}}
\tabletypesize{\scriptsize}
\tablehead{
\colhead{Object} &
\colhead{$z$} &
\colhead{$D_L$} &
\colhead{log $M_*$} &
\colhead{$Z$} & 
\colhead{log $\tau_{9.7}$}  & 
\colhead{log $ L_{\rm [Ne\ II] +[Ne\ III] } $} &
\colhead{$ {\rm [Ne\ III]} / {\rm [Ne\ II]} $} &
\colhead{$ f_{\rm +}$}  &
\colhead{$ f_{\rm 2+}$} &
\colhead{log SFR } &
\colhead{Notes} \\
\colhead{} &
\colhead{} &
\colhead{(Mpc)} &
\colhead{($M_\odot$)} &
\colhead{12+log (O/H)} &  
\colhead{} &
\colhead{(erg s$^{-1}$)} &
\colhead{} &
\colhead{} &
\colhead{} &
\colhead{($M_{\odot}\rm\,yr^{-1}$)} &
\colhead{} 
\\ 
\colhead{(1)} &
\colhead{(2)} &
\colhead{(3)} &
\colhead{(4)} &
\colhead{(5)} &
\colhead{(6)} &
\colhead{(7)} &
\colhead{(8)} &
\colhead{(9)} &
\colhead{(10)} &
\colhead{(11)} &
\colhead{(12)}
}
\startdata
%### obj, redshift, Dlum, log M*, Z, tau9.7, log (Ne II+ Ne III),Ne III/Ne II, f1, f2, SFR,  source type ###%
CG~0752$^{\ast}$                    & 0.021  &   100  &           $ 9.88_{-0.21}^{+0.21} $  &                           $ 8.50 $  &          $ -0.57_{-0.10}^{+0.10} $  &          $ 41.34_{-0.01}^{+0.01} $  &           $ 0.47_{-0.02}^{+0.02} $  &           $ 0.72_{-0.01}^{+0.01} $  &           $ 0.21_{-0.01}^{+0.01} $  &           $ 1.13_{-0.01}^{+0.01} $  &   BCD
                                 \\
Haro~11$^{\ast}$                    & 0.021  &    90  &          $ 10.24_{-0.20}^{+0.20} $  &                           $ 7.69 $  &          $ -0.43_{-0.10}^{+0.10} $  &          $ 42.12_{-0.01}^{+0.01} $  &           $ 3.17_{-0.13}^{+0.13} $  &           $ 0.32_{-0.01}^{+0.01} $  &           $ 0.63_{-0.01}^{+0.01} $  &           $ 2.61_{-0.01}^{+0.01} $  &   BCD
                                 \\
II~Zw~40$^{\ast}$                   & 0.003  &    10  &           $ 8.12_{-0.21}^{+0.21} $  &                           $ 7.92 $  &          $ -0.79_{-0.03}^{+0.03} $  &          $ 40.15_{-0.01}^{+0.01} $  &          $ 19.50_{-1.24}^{+1.24} $  &           $ 0.07_{-0.01}^{+0.01} $  &           $ 0.92_{-0.01}^{+0.01} $  &           $ 0.35_{-0.01}^{+0.01} $  &   BCD
                                 \\
I~Zw~18                    & 0.002  &    18  &           $ 6.12_{-0.09}^{+0.10} $  &           $ 7.18_{-0.01}^{+0.01} $  &          $ -0.18_{-0.47}^{+0.47} $  &          $ 38.10_{-0.11}^{+0.11} $  &                        $>  10.00 $  &                           $ 0.02 $  &                           $ 1.00 $  &          $ -0.97_{-0.11}^{+0.11} $  &   BCD
                                 \\
KUG~1013+381$^{\ast}$               & 0.004  &    23  &           $ 7.52_{-0.22}^{+0.22} $  &           $ 7.57_{-0.01}^{+0.01} $  &          $ -0.12_{-0.07}^{+0.07} $  &          $ 38.84_{-0.06}^{+0.06} $  &                         $>  4.31 $  &                           $ 0.02 $  &                           $ 1.00 $  &          $ -0.63_{-0.06}^{+0.06} $  &   BCD
                                 \\
Mrk~1450                   & 0.003  &    20  &           $ 6.86_{-0.07}^{+0.05} $  &           $ 7.98_{-0.01}^{+0.01} $  &                         $< -0.43 $  &          $ 39.70_{-0.04}^{+0.04} $  &          $ 10.60_{-1.79}^{+1.79} $  &           $ 0.12_{-0.02}^{+0.02} $  &           $ 0.86_{-0.02}^{+0.02} $  &          $ -0.15_{-0.04}^{+0.04} $  &   BCD
                                 \\
Mrk~1499$^{\ast}$                   & 0.009  &    41  &           $ 8.60_{-0.21}^{+0.21} $  &                           $ 8.17 $  &                         $< -1.98 $  &          $ 40.04_{-0.01}^{+0.01} $  &           $ 2.41_{-0.11}^{+0.11} $  &           $ 0.38_{-0.01}^{+0.01} $  &           $ 0.57_{-0.01}^{+0.01} $  &           $ 0.08_{-0.01}^{+0.01} $  &   BCD
                                 \\
Mrk~475$^{\ast}$                    & 0.002  &     9  &           $ 7.15_{-0.22}^{+0.22} $  &           $ 7.93_{-0.01}^{+0.01} $  &                         $< -0.86 $  &          $ 38.37_{-0.03}^{+0.03} $  &                         $>  9.46 $  &                           $ 0.02 $  &                           $ 1.00 $  &          $ -1.46_{-0.03}^{+0.03} $  &   BCD
                                 \\
Mrk~487$^{\ast}$                    & 0.002  &    15  &           $ 8.12_{-0.21}^{+0.21} $  &           $ 8.06_{-0.04}^{+0.04} $  &                         $< -1.40 $  &          $ 38.98_{-0.02}^{+0.02} $  &           $ 5.51_{-0.83}^{+0.83} $  &           $ 0.22_{-0.03}^{+0.03} $  &           $ 0.75_{-0.03}^{+0.03} $  &          $ -0.91_{-0.05}^{+0.05} $  &   BCD                                 \\
IRAS~00188$-$0856$^{\ast}$          & 0.128  &   619  &          $ 11.02_{-0.20}^{+0.20} $  &           $ 8.89_{-0.12}^{+0.12} $  &           $ 0.64_{-0.02}^{+0.02} $  &          $ 43.45_{-0.05}^{+0.05} $  &           $ 0.14_{-0.04}^{+0.04} $  &           $ 0.88_{-0.02}^{+0.02} $  &           $ 0.07_{-0.02}^{+0.02} $  &           $ 2.89_{-0.13}^{+0.13} $  & ULIRG
                                 \\
IRAS~06035$-$7102$^{\ast}$          & 0.079  &   369  &          $ 11.15_{-0.20}^{+0.20} $  &           $ 8.88_{-0.11}^{+0.11} $  &           $ 0.37_{-0.03}^{+0.03} $  &          $ 42.66_{-0.03}^{+0.03} $  &           $ 0.29_{-0.04}^{+0.04} $  &           $ 0.79_{-0.02}^{+0.02} $  &           $ 0.14_{-0.02}^{+0.02} $  &           $ 2.09_{-0.12}^{+0.12} $  & ULIRG
                                 \\
IRAS~10565+2448$^{\ast}$            & 0.043  &   196  &          $ 10.99_{-0.20}^{+0.20} $  &           $ 8.89_{-0.12}^{+0.12} $  &           $ 0.25_{-0.01}^{+0.01} $  &          $ 42.91_{-0.01}^{+0.01} $  &           $ 0.12_{-0.01}^{+0.01} $  &           $ 0.88_{-0.01}^{+0.01} $  &           $ 0.06_{-0.01}^{+0.01} $  &           $ 2.35_{-0.12}^{+0.12} $  & ULIRG
                                 \\
IRAS~11095$-$0238$^{\ast}$          & 0.107  &   510  &          $ 10.37_{-0.20}^{+0.20} $  &           $ 8.85_{-0.15}^{+0.15} $  &           $ 0.77_{-0.01}^{+0.01} $  &          $ 43.79_{-0.02}^{+0.02} $  &           $ 0.30_{-0.03}^{+0.03} $  &           $ 0.79_{-0.01}^{+0.01} $  &           $ 0.15_{-0.01}^{+0.01} $  &           $ 3.25_{-0.15}^{+0.15} $  & ULIRG
                                 \\
IRAS~14070+0525$^{\ast}$            & 0.264  &  1380  &          $ 11.13_{-0.21}^{+0.21} $  &           $ 8.89_{-0.12}^{+0.12} $  &           $ 0.76_{-0.01}^{+0.01} $  &          $ 43.75_{-0.08}^{+0.08} $  &                         $<  0.68 $  &                           $ 0.94 $  &                           $ 0.02 $  &           $ 3.20_{-0.14}^{+0.14} $  & ULIRG
                                 \\
IRAS~15462$-$0450$^{\ast}$          & 0.100  &   474  &          $ 10.86_{-0.20}^{+0.20} $  &           $ 8.89_{-0.13}^{+0.13} $  &          $ -0.80_{-0.82}^{+0.82} $  &          $ 42.52_{-0.06}^{+0.06} $  &           $ 0.38_{-0.09}^{+0.09} $  &           $ 0.76_{-0.04}^{+0.04} $  &           $ 0.18_{-0.04}^{+0.04} $  &           $ 1.93_{-0.15}^{+0.15} $  & ULIRG
                                 \\
IRAS~16090$-$0139$^{\ast}$          & 0.134  &   650  &          $ 11.19_{-0.20}^{+0.20} $  &           $ 8.88_{-0.11}^{+0.11} $  &           $ 0.69_{-0.01}^{+0.01} $  &          $ 43.83_{-0.01}^{+0.01} $  &           $ 0.18_{-0.02}^{+0.02} $  &           $ 0.85_{-0.01}^{+0.01} $  &           $ 0.10_{-0.01}^{+0.01} $  &           $ 3.28_{-0.11}^{+0.11} $  & ULIRG
                                 \\
IRAS~17208$-$0014$^{\ast}$          & 0.043  &   196  &          $ 10.89_{-0.20}^{+0.20} $  &           $ 8.89_{-0.13}^{+0.13} $  &           $ 0.53_{-0.01}^{+0.01} $  &          $ 43.13_{-0.01}^{+0.01} $  &           $ 0.21_{-0.01}^{+0.01} $  &           $ 0.83_{-0.01}^{+0.01} $  &           $ 0.11_{-0.01}^{+0.01} $  &           $ 2.56_{-0.13}^{+0.13} $  & ULIRG
                                 \\
IRAS~20087$-$0308$^{\ast}$          & 0.106  &   505  &          $ 11.10_{-0.20}^{+0.20} $  &           $ 8.89_{-0.12}^{+0.12} $  &           $ 0.45_{-0.01}^{+0.01} $  &          $ 43.31_{-0.01}^{+0.01} $  &           $ 0.12_{-0.01}^{+0.01} $  &           $ 0.88_{-0.01}^{+0.01} $  &           $ 0.06_{-0.01}^{+0.01} $  &           $ 2.76_{-0.12}^{+0.12} $  & ULIRG
                                 \\
2MASS~J10365355+5754426    & 0.102  &   485  &          $ 10.29_{-0.11}^{+0.09} $  &           $ 8.84_{-0.15}^{+0.15} $  &                         $< -2.01 $  &          $ 41.42_{-0.05}^{+0.05} $  &           $ 0.49_{-0.09}^{+0.09} $  &           $ 0.71_{-0.03}^{+0.03} $  &           $ 0.23_{-0.03}^{+0.03} $  &           $ 0.87_{-0.16}^{+0.16} $  &   SFG
                                 \\
2MASX~J10372318+5731144    & 0.072  &   335  &          $ 10.67_{-0.10}^{+0.08} $  &           $ 8.89_{-0.14}^{+0.14} $  &                         $< -2.52 $  &          $ 41.81_{-0.01}^{+0.01} $  &           $ 0.19_{-0.01}^{+0.01} $  &           $ 0.84_{-0.01}^{+0.01} $  &           $ 0.10_{-0.01}^{+0.01} $  &           $ 1.24_{-0.14}^{+0.14} $  &   SFG
                                 \\
2MASX~J10413896+5835150    & 0.115  &   551  &          $ 11.09_{-0.12}^{+0.09} $  &           $ 8.89_{-0.12}^{+0.12} $  &          $ -0.73_{-0.04}^{+0.04} $  &          $ 41.88_{-0.04}^{+0.04} $  &                         $<  0.15 $  &                           $ 0.94 $  &                           $ 0.02 $  &           $ 1.33_{-0.12}^{+0.12} $  &   SFG
                                 \\
2MASX~J10430576+5841518    & 0.119  &   572  &          $ 10.57_{-0.11}^{+0.08} $  &           $ 8.88_{-0.14}^{+0.14} $  &                         $< -2.18 $  &          $ 41.68_{-0.06}^{+0.06} $  &                         $<  0.28 $  &                           $ 0.94 $  &                           $ 0.02 $  &           $ 1.14_{-0.15}^{+0.15} $  &   SFG
                                 \\
2MASX~J10433909+5802509    & 0.133  &   645  &          $ 10.65_{-0.11}^{+0.08} $  &           $ 8.88_{-0.14}^{+0.14} $  &          $ -1.08_{-0.12}^{+0.12} $  &          $ 42.07_{-0.03}^{+0.03} $  &                         $<  0.38 $  &                           $ 0.94 $  &                           $ 0.02 $  &           $ 1.53_{-0.14}^{+0.14} $  &   SFG
                                 \\
2MASX~J10440070+5845370    & 0.073  &   340  &          $ 10.84_{-0.12}^{+0.11} $  &           $ 8.90_{-0.13}^{+0.13} $  &          $ -0.20_{-0.01}^{+0.01} $  &          $ 41.74_{-0.01}^{+0.01} $  &           $ 0.12_{-0.02}^{+0.02} $  &           $ 0.88_{-0.01}^{+0.01} $  &           $ 0.06_{-0.01}^{+0.01} $  &           $ 1.18_{-0.13}^{+0.13} $  &   SFG
                                 \\
2MASX~J10453036+5812322    & 0.118  &   567  &          $ 11.24_{-0.11}^{+0.09} $  &           $ 8.88_{-0.11}^{+0.11} $  &          $ -1.37_{-0.11}^{+0.11} $  &          $ 41.95_{-0.03}^{+0.03} $  &           $ 0.19_{-0.03}^{+0.03} $  &           $ 0.84_{-0.02}^{+0.02} $  &           $ 0.10_{-0.01}^{+0.01} $  &           $ 1.39_{-0.11}^{+0.11} $  &   SFG
                                 \\
2MASX~J10470453+5620253    & 0.047  &   215  &          $ 10.43_{-0.11}^{+0.08} $  &           $ 8.86_{-0.15}^{+0.15} $  &          $ -0.27_{-0.01}^{+0.01} $  &          $ 41.62_{-0.01}^{+0.01} $  &           $ 0.12_{-0.01}^{+0.01} $  &           $ 0.88_{-0.01}^{+0.01} $  &           $ 0.06_{-0.01}^{+0.01} $  &           $ 1.09_{-0.15}^{+0.15} $  &   SFG
                                 \\
2MASX~J10471339+5810393    & 0.061  &   282  &          $ 10.05_{-0.11}^{+0.09} $  &           $ 8.80_{-0.15}^{+0.15} $  &          $ -2.77_{-3.45}^{+3.45} $  &          $ 40.98_{-0.04}^{+0.04} $  &           $ 0.20_{-0.10}^{+0.10} $  &           $ 0.84_{-0.05}^{+0.05} $  &           $ 0.10_{-0.05}^{+0.05} $  &           $ 0.51_{-0.16}^{+0.16} $  &   SFG
                                 \\
\enddata
\tablecomments{
Col. (1) Object name; asterisk ($\ast$) indicates an object whose stellar mass is calculated in the current work.
Col. (2) Redshift.
Col. (3) Luminosity distance.
Col. (4) Stellar mass. 
Col. (5) Metallicity. We collect the metallicity measurements of Haro~11, NGC~1140, II~Zw~40, UGC~4274,  CG~0752, and Mrk~1499 from Wu \etal (2006); Mrk~1450, UM~461, SBS~0335$-$052E, I~Zw~18, Mrk~487,  SBS~1030+583, SBS~1415+437, SBS~1159+545, and Mrk~475 from Izotov \& Thuan (1999); KUG~1013+381 from Izotov et al. (2012); SDSS~J0825+3532 from Berg \etal (2016); and Tol~65 from Kobulnicky \& Skillman (1996). 
Col. (6) Dust extinction at 9.7$\mum$.
Col. (7) Luminosity of [Ne II]+[Ne III]. %12.81$\mum$ 15.56$\mum$
Col. (8) Luminosity ratio of [Ne III]/[Ne II] .
Col. (9) Fractional abundance of neon in [Ne II] state.
Col. (10) Fractional abundance of neon in [Ne III] state.
Col. (11) Star formation rate. 
Col. (12) Galaxy type classified based on bolometric luminosity: 
star-forming galaxy (SFG);
blue compact dwarf (BCD); 
ultraluminous infrared galaxy (ULIRG).
The complete table is available in the online journal.
}
\end{deluxetable}
\end{longrotatetable}
%%%%%%%%%%%%%%%%%%%%%%%%%%%%%%%%%%%%%%%%%%%%%%

%\input{table_pah_long.tex}
%%%%%%%%%%%%%%%%%%%%%%%%%%%%%%%%%%%%%%%%%%%%%%
%Edited LCH,. April 17, 2019
\begin{longrotatetable}
\begin{deluxetable}{l c c c c c c}%l }
\tablecaption{PAH Luminosity in Different Bands  \label{tab:lumpah}}
\tabletypesize{\scriptsize}
\tablehead{
\colhead{Object} &
\colhead{log $ L_{\rm PAH}(6.2\mum)$}  &
\colhead{log $ L_{\rm PAH} (7.7\mum)$}  &
\colhead{log $ L_{\rm PAH}(8.6\mum)$}  &
\colhead{log $ L_{\rm PAH}(11.3\mum)$}  &
\colhead{log $ L_{\rm PAH}(5-15\mum)$} &
\colhead{Notes}  \\
\colhead{} &
\colhead{(erg s$^{-1}$)} &
\colhead{(erg s$^{-1}$)} &
\colhead{(erg s$^{-1}$)} &
\colhead{(erg s$^{-1}$)} &
\colhead{(erg s$^{-1}$)} &
\colhead{}     \\ 
\colhead{(1)} &
\colhead{(2)} &
\colhead{(3)} &
\colhead{(4)} &
\colhead{(5)} &
\colhead{(6)} &
\colhead{(7)} 
}
\startdata
%### obj, log pah6.2, log pah7.7, log pah8.6, log pah11.2,log pah515, source type ###
CG~0752                    &          $ 41.60_{-0.01}^{+0.01} $  &          $ 42.40_{-0.01}^{+0.01} $  &          $ 41.72_{-0.03}^{+0.03} $  &          $ 41.83_{-0.02}^{+0.02} $  &          $ 42.88_{-0.01}^{+0.01} $  &     BCD \\
Haro~11                    &          $ 41.78_{-0.02}^{+0.02} $  &          $ 42.74_{-0.03}^{+0.03} $  &          $ 42.00_{-0.13}^{+0.13} $  &          $ 42.09_{-0.10}^{+0.10} $  &          $ 43.19_{-0.02}^{+0.02} $  &     BCD \\
II~Zw~40                   &          $ 38.96_{-0.04}^{+0.04} $  &          $ 39.84_{-0.04}^{+0.04} $  &                         $< 39.20 $  &          $ 39.60_{-0.06}^{+0.06} $  &          $ 40.38_{-0.04}^{+0.04} $  &     BCD \\
I~Zw~18                    &                         $< 39.17 $  &                         $< 39.46 $  &                         $< 39.04 $  &                         $< 38.83 $  &                         $< 40.33 $  &     BCD \\
KUG~1013+381               &                         $< 38.14 $  &                         $< 38.88 $  &                         $< 38.66 $  &                         $< 37.91 $  &                         $< 39.71 $  &     BCD \\
Mrk~1450                   &                         $< 38.93 $  &                         $< 39.55 $  &                         $< 39.28 $  &                         $< 39.31 $  &                         $< 39.96 $  &     BCD \\
Mrk~1499                   &          $ 40.11_{-0.04}^{+0.04} $  &          $ 40.89_{-0.04}^{+0.04} $  &          $ 40.29_{-0.05}^{+0.05} $  &          $ 40.45_{-0.02}^{+0.02} $  &          $ 41.38_{-0.02}^{+0.02} $  &     BCD \\
Mrk~475                    &          $ 37.73_{-0.07}^{+0.07} $  &          $ 38.21_{-0.06}^{+0.06} $  &                         $< 37.72 $  &          $ 37.94_{-0.09}^{+0.09} $  &          $ 38.76_{-0.06}^{+0.06} $  &     BCD \\
Mrk~487                    &                         $< 38.57 $  &                         $< 39.10 $  &                          $<$ 39.10  &          $ 39.18_{-0.08}^{+0.08} $  &                         $< 39.63 $  &     BCD \\
NGC~1140                   &          $ 40.19_{-0.01}^{+0.01} $  &          $ 40.94_{-0.02}^{+0.02} $  &          $ 40.16_{-0.04}^{+0.04} $  &          $ 40.51_{-0.02}^{+0.02} $  &          $ 41.40_{-0.01}^{+0.01} $  &     BCD \\
IRAS~00188$-$0856          &          $ 43.44_{-0.14}^{+0.14} $  &          $ 44.56_{-0.08}^{+0.08} $  &          $ 44.32_{-0.12}^{+0.12} $  &          $ 44.22_{-0.08}^{+0.08} $  &          $ 45.17_{-0.05}^{+0.05} $  &   ULIRG \\
IRAS~06035$-$7102          &          $ 42.96_{-0.14}^{+0.14} $  &          $ 44.14_{-0.06}^{+0.06} $  &          $ 43.57_{-0.14}^{+0.14} $  &          $ 43.54_{-0.10}^{+0.10} $  &          $ 44.55_{-0.04}^{+0.04} $  &   ULIRG \\
IRAS~10565+2448            &          $ 43.24_{-0.01}^{+0.01} $  &          $ 43.96_{-0.01}^{+0.01} $  &          $ 43.43_{-0.01}^{+0.01} $  &          $ 43.51_{-0.01}^{+0.01} $  &          $ 44.51_{-0.01}^{+0.01} $  &   ULIRG \\
IRAS~11095$-$0238          &          $ 43.66_{-0.14}^{+0.14} $  &          $ 45.00_{-0.04}^{+0.04} $  &          $ 44.17_{-0.14}^{+0.14} $  &          $ 44.24_{-0.12}^{+0.12} $  &          $ 45.54_{-0.03}^{+0.03} $  &   ULIRG \\
IRAS~14070+0525            &          $ 44.20_{-0.14}^{+0.14} $  &          $ 45.39_{-0.06}^{+0.06} $  &          $ 44.76_{-0.15}^{+0.15} $  &          $ 44.79_{-0.12}^{+0.12} $  &          $ 45.87_{-0.04}^{+0.04} $  &   ULIRG \\
IRAS~15462$-$0450          &          $ 42.84_{-0.14}^{+0.14} $  &          $ 43.59_{-0.14}^{+0.14} $  &          $ 43.38_{-0.14}^{+0.14} $  &          $ 43.12_{-0.15}^{+0.15} $  &          $ 44.01_{-0.13}^{+0.13} $  &   ULIRG \\
IRAS~16090$-$0139          &          $ 43.73_{-0.05}^{+0.05} $  &          $ 44.93_{-0.02}^{+0.02} $  &          $ 44.08_{-0.09}^{+0.09} $  &          $ 44.53_{-0.02}^{+0.02} $  &          $ 45.49_{-0.01}^{+0.01} $  &   ULIRG \\
IRAS~17208$-$0014          &          $ 43.56_{-0.01}^{+0.01} $  &          $ 44.37_{-0.01}^{+0.01} $  &          $ 43.71_{-0.01}^{+0.01} $  &          $ 43.91_{-0.01}^{+0.01} $  &          $ 44.91_{-0.01}^{+0.01} $  &   ULIRG \\
IRAS~20087$-$0308          &          $ 43.58_{-0.01}^{+0.01} $  &          $ 44.52_{-0.01}^{+0.01} $  &          $ 43.85_{-0.01}^{+0.01} $  &          $ 43.98_{-0.01}^{+0.01} $  &          $ 45.03_{-0.01}^{+0.01} $  &   ULIRG \\
IRAS~20100$-$4156          &          $ 43.62_{-0.04}^{+0.04} $  &          $ 44.94_{-0.01}^{+0.01} $  &          $ 44.29_{-0.02}^{+0.02} $  &          $ 44.45_{-0.02}^{+0.02} $  &          $ 45.52_{-0.01}^{+0.01} $  &   ULIRG \\
%IRAS~20551$-$4250          &          $ 43.64_{-0.01}^{+0.01} $  &          $ 44.57_{-0.01}^{+0.01} $  &          $ 43.83_{-0.03}^{+0.03} $  &          $ 44.05_{-0.02}^{+0.02} $  &          $ 45.14_{-0.01}^{+0.01} $  &   ULIRG \\
%IRAS~22491$-$1808          &          $ 43.12_{-0.01}^{+0.01} $  &          $ 43.92_{-0.01}^{+0.01} $  &          $ 43.25_{-0.03}^{+0.03} $  &          $ 43.39_{-0.02}^{+0.02} $  &          $ 44.44_{-0.01}^{+0.01} $  &   ULIRG \\
%IRAS~23253$-$5415          &          $ 43.00_{-0.08}^{+0.08} $  &          $ 43.96_{-0.06}^{+0.06} $  &          $ 43.42_{-0.14}^{+0.14} $  &          $ 43.46_{-0.09}^{+0.09} $  &          $ 44.46_{-0.03}^{+0.03} $  &   ULIRG \\
2MASS~J10365355+5754426    &          $ 41.97_{-0.01}^{+0.01} $  &          $ 42.78_{-0.01}^{+0.01} $  &          $ 42.14_{-0.01}^{+0.01} $  &          $ 42.25_{-0.01}^{+0.01} $  &          $ 43.25_{-0.01}^{+0.01} $  &     SFG \\
2MASX~J10372318+5731144    &          $ 41.96_{-0.01}^{+0.01} $  &          $ 43.05_{-0.01}^{+0.01} $  &          $ 42.47_{-0.01}^{+0.01} $  &          $ 42.58_{-0.01}^{+0.01} $  &          $ 43.52_{-0.01}^{+0.01} $  &     SFG \\
2MASX~J10413896+5835150    &          $ 42.51_{-0.01}^{+0.01} $  &          $ 43.36_{-0.01}^{+0.01} $  &          $ 42.70_{-0.01}^{+0.01} $  &          $ 42.81_{-0.01}^{+0.01} $  &          $ 43.86_{-0.01}^{+0.01} $  &     SFG \\
2MASX~J10430576+5841518    &          $ 41.29_{-0.09}^{+0.09} $  &          $ 42.85_{-0.01}^{+0.01} $  &          $ 42.28_{-0.02}^{+0.02} $  &          $ 42.44_{-0.01}^{+0.01} $  &          $ 43.32_{-0.01}^{+0.01} $  &     SFG \\
2MASX~J10433909+5802509    &          $ 42.42_{-0.01}^{+0.01} $  &          $ 43.12_{-0.01}^{+0.01} $  &          $ 42.33_{-0.02}^{+0.02} $  &          $ 42.58_{-0.01}^{+0.01} $  &          $ 43.62_{-0.01}^{+0.01} $  &     SFG \\
2MASX~J10440070+5845370    &          $ 42.21_{-0.01}^{+0.01} $  &          $ 43.05_{-0.01}^{+0.01} $  &          $ 42.46_{-0.01}^{+0.01} $  &          $ 42.49_{-0.01}^{+0.01} $  &          $ 43.54_{-0.01}^{+0.01} $  &     SFG \\
2MASX~J10453036+5812322    &          $ 42.48_{-0.01}^{+0.01} $  &          $ 43.35_{-0.01}^{+0.01} $  &          $ 42.65_{-0.01}^{+0.01} $  &          $ 42.81_{-0.01}^{+0.01} $  &          $ 43.83_{-0.01}^{+0.01} $  &     SFG \\
2MASX~J10470453+5620253    &          $ 42.15_{-0.01}^{+0.01} $  &          $ 42.98_{-0.01}^{+0.01} $  &          $ 42.41_{-0.01}^{+0.01} $  &          $ 42.44_{-0.01}^{+0.01} $  &          $ 43.48_{-0.01}^{+0.01} $  &     SFG \\
2MASX~J10471339+5810393    &          $ 41.52_{-0.01}^{+0.01} $  &          $ 42.36_{-0.01}^{+0.01} $  &          $ 41.68_{-0.01}^{+0.01} $  &          $ 41.78_{-0.01}^{+0.01} $  &          $ 42.81_{-0.01}^{+0.01} $  &     SFG \\
2MASX~J10474253+5618500    &          $ 42.44_{-0.01}^{+0.01} $  &          $ 43.29_{-0.01}^{+0.01} $  &          $ 42.64_{-0.01}^{+0.01} $  &          $ 42.78_{-0.01}^{+0.01} $  &          $ 43.78_{-0.01}^{+0.01} $  &     SFG \\
\enddata
\tablecomments{
Col. (1) Object name.
Col. (2) Luminosity of  6.2$\mum$ PAH feature.
Col. (3) Luminosity of  7.7$\mum$ PAH feature.
Col. (4) Luminosity of  8.6$\mum$ PAH feature.
Col. (5) Luminosity of  11.3$\mum$ PAH feature.
Col. (6) Luminosity of  PAH in $5-15$$\mum$ band.
%Col. (7) Luminosity of  PAH in $5-30$$\mum$ band.
Col. (7)  Galaxy type classified based on bolometric luminosity:
star-forming galaxy (SFG);
blue compact dwarf (BCD);
ultraluminous infrared galaxy (ULIRG).
The complete table is available in the online journal.
}
\end{deluxetable}
\end{longrotatetable}
%%%%%%%%%%%%%%%%%%%%%%%%%%%%%%%%%%%%%%%%%%%%%%

\section{SFR Calibration \label{sec:result}}

% Axis labels, here and throughout, the variables should not be in boldface
% I don't know how to do this so far ... 

\begin{figure}%[ht]
\begin{center}
$
\begin{array}{ccc}
\includegraphics[height=0.3\textwidth]{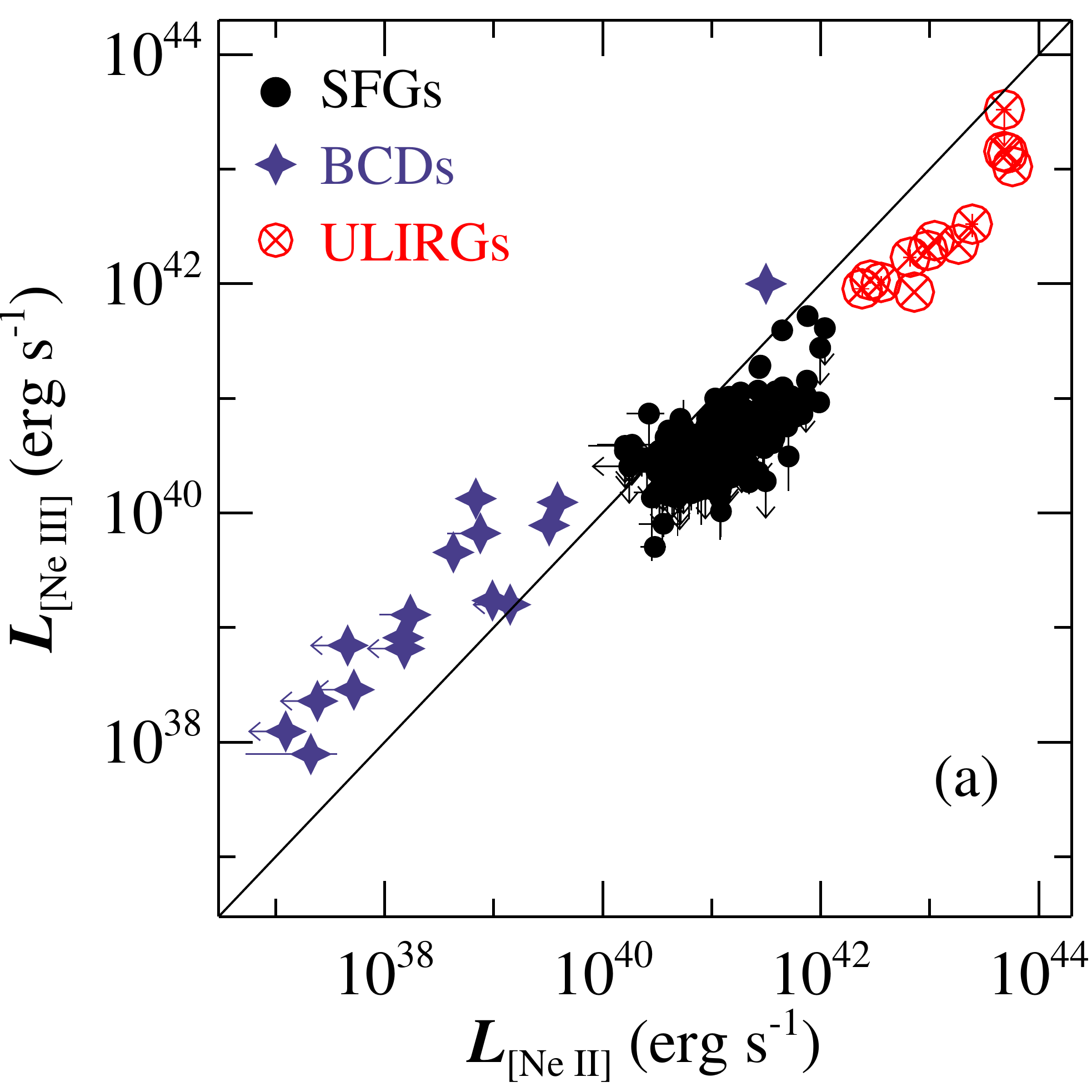} &
\includegraphics[height=0.3\textwidth]{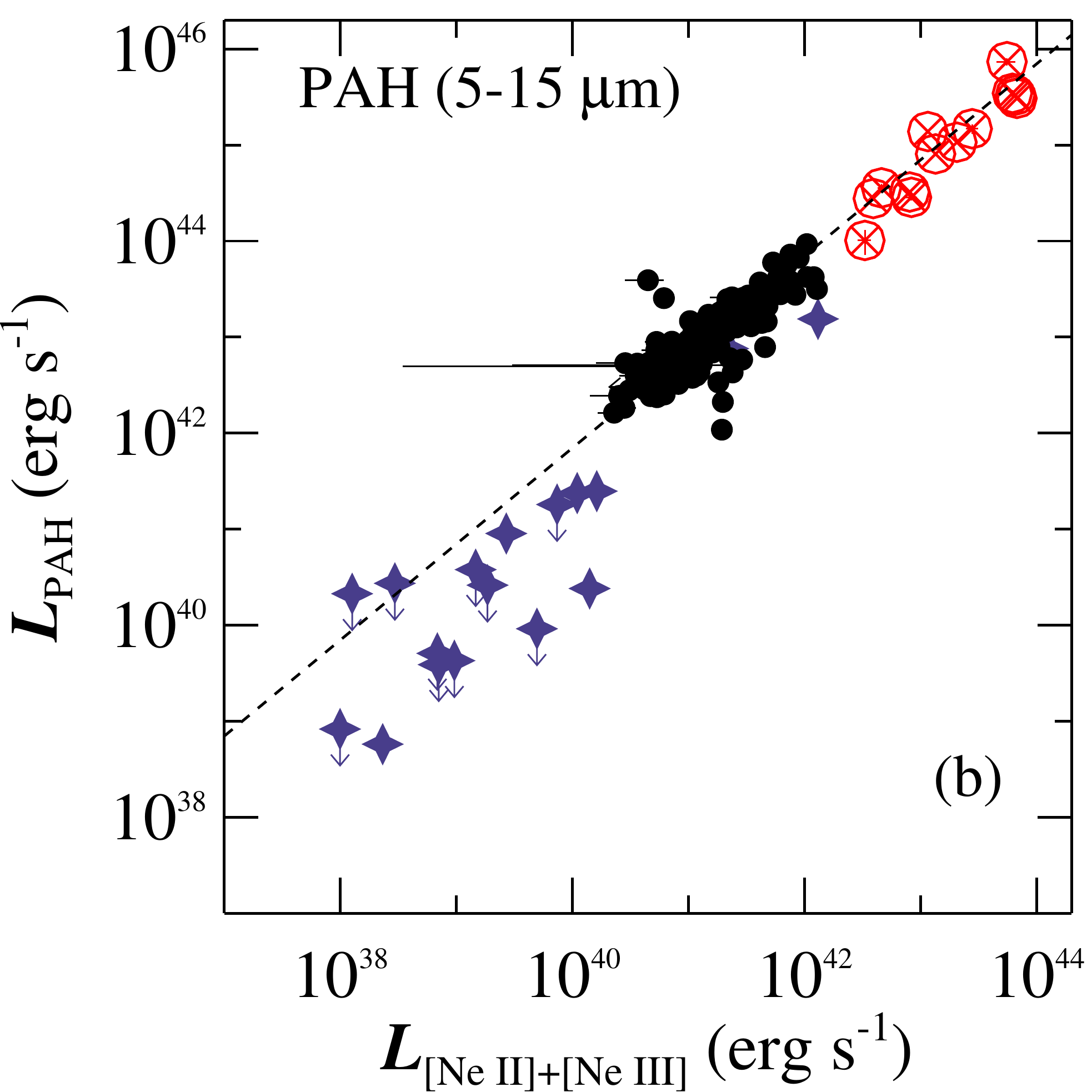} &
\includegraphics[height=0.3\textwidth]{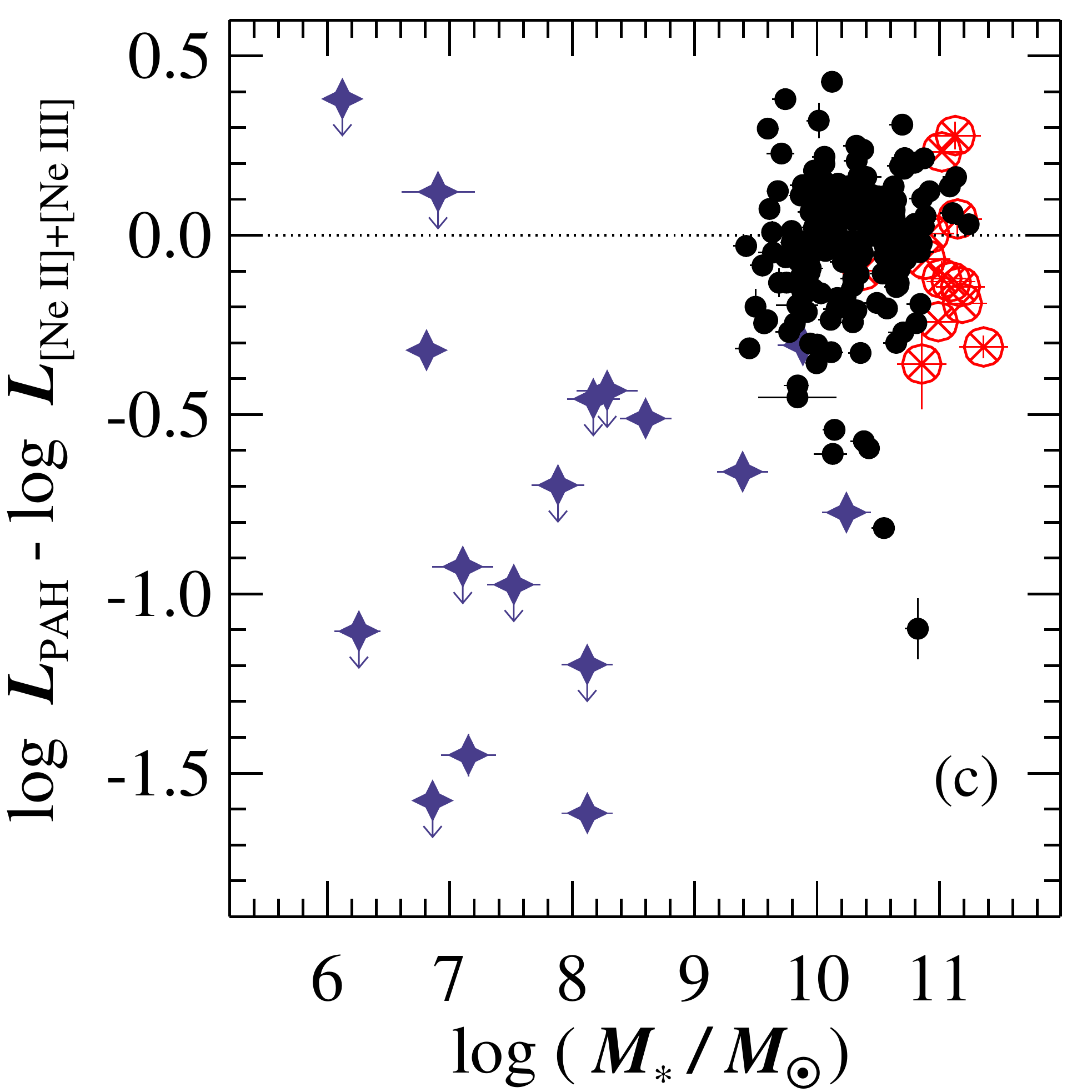}
\end{array}
$
\caption{ 
(a) Relationship between [Ne~II] 12.8$\mum$ and [Ne~III] 15.6$\mum$ luminosity; the solid line shows the 1:1 relation.  (b) Correlation between the luminosity of [Ne~II]+[Ne~III] ($L_{\rm [Ne~{\sc II}]+ [Ne~{\sc III}]}$) and total PAH emission ($L_{\mathrm {PAH}}$) in the wavelength range $5-15\mum$. The dashed line represents the median value of $L_{\rm [Ne~{\sc II}]+ [Ne~{\sc III}]}/L_{\mathrm{PAH}}$ for normal star-forming galaxies and starburst-dominated ULIRGs.  (c) The residuals of panel (b), $\log L_{\mathrm {PAH}} - \log L_{\rm [Ne~{\sc II}]+ [Ne~{\sc III}]}$, are plotted as a function of stellar mass. 
\label{fig:L_neon_pah}}
\end{center}
\end{figure}

\subsection{SFR Based on PAH Emission \label{subsec:pahsfr}}

Our PAH-based SFR estimator is directly calibrated against SFRs for our calibration sample determined from the [Ne~II] 12.8$\mum$ and [Ne~III] 15.6$\mum$ lines, which for our objects span $\sim 6-7$ orders of magnitude in luminosity.  As discussed in Ho \& Keto (2007; see also Thornley \etal 2000), the combined luminosity of [Ne~II] + [Ne~III] traces the ionizing luminosity of massive stars, and hence can be used as an effective estimator of the SFR.  For the massive star-forming and starburst galaxies in our sample, the median value of [Ne~III]/[Ne~II] $\approx 0.3$, while for the low-mass BCDs the ratio is $\sim 20$ times higher, at [Ne~III]/[Ne~II] $\approx 5.5$ (Figure~\ref{fig:L_neon_pah}a).  The enhancement in the ionization of the line-emitting gas in dwarf galaxies arises from the harder ionizing radiation of low-metallicity systems (Madden \etal 2006).  Figure~\ref{fig:L_neon_pah}b demonstrates that the PAH luminosity integrated over $5-15 \mum$ ($L_\mathrm{PAH}$) correlates closely with the total neon emission ($L_{\rm [Ne~{\sc II}]+ [Ne~{\sc III}]}$) when $L_\mathrm{PAH} \gtrsim 10^{42}\,\rm erg\,s^{-1}$.  This regime is mostly occupied by massive galaxies.  By contrast, BCDs, which in our sample have $L_\mathrm{PAH} \lesssim 10^{42} \,\rm erg\,s^{-1}$, uniformly exhibit a deficit in PAH emission relative to neon.  As discussed in Section 5.1, the main physical driver for the PAH deficit in BCDs is metallicity, or, indirectly, stellar mass, given the relationship between the two (e.g., Tremonti et al. 2004).  The onset of the PAH deficit occurs below a stellar mass of $M_{\ast} \approx 10^{9}\,M_{\odot}$, and the magnitude of the deficit decreases systematically with increasing $M_{\ast}$, albeit with significant scatter (Figure~\ref{fig:L_neon_pah}c).

Zhuang et al. (2019) updated the neon-based SFR relation of Ho \& Keto (2007) to explicitly include the effect of metallicity:

\begin{equation}
{\rm SFR} \ (M_{\odot}\ {\rm yr^{-1}}) = 4.34\times10^{-41}  \  
\left( Z_{\odot}/Z \right) \ 
\left[ { { L_{\rm [Ne~{\sc II}]+ [Ne~{\sc III}]} \ ({\rm erg} \ \ {\rm s}^{-1}) }\over{ f_{+} + 1.67 f_{+2} } } \right],
\label{eq:ho_sfr}
\end{equation}

\noindent
where $Z$ is the gas-phase metallicity of the galaxy, and $f_{+}$ and $f_{+2}$ are the fractional abundances of singly and doubly ionized neon. The fractional abundances of ionized neon depend on a number of factors, as described in Zhuang \etal (2019), who performed photoionization calculations over a wide range of metallicities and star formation histories, assuming a Salpeter (1955) stellar initial mass function.  Zhuang et al. (2019; their Equation 5) give an explicit fitting function to estimate $f_{+}$ and $f_{+2}$ based on the observed [Ne~III]/[Ne~II] ratio. Uncertainties are estimated from bootstrap resampling.  The derived fractional abundances populate well the full range of values predicted from the photoionization models.  For galaxies with [Ne III] undetected, we adopt the values of $f_{+}$ and $f_{+2}$ in the limit [Ne~III]/[Ne~II] $\rightarrow -\infty$, whereas we adopt the values for [Ne~III]/[Ne~II] $\rightarrow \infty$ when [Ne~II] is undetected.  For galaxies with neither line detected, we take the peak values of $f_{+}=0.75$ and $f_{+2}=0.15$ (Ho \& Keto 2007). 

We estimate the metallicities for the massive galaxies from the mass-metallicity relation of Kewley \& Ellison (2008), who derived metallicities based on a combination of the N2O2 (Kewley \& Dopita 2002) and N2 (Pettini \& Pagel 2004) methods.  The N2O2 method yields higher absolute metallicities, while the N2 approach gives lower values.  We take the average of the two as the final estimate and use the results from the N2O2 and N2 methods to bracket the upper and lower limits, respectively.  The mass-metallicity relation has significant scatter and overestimates the metallicity for dwarf galaxies with $M_{\ast} < 10^{7.5} M_{\odot}$ (\eg Jimmy \etal 2015; Yates \etal 2019); thus, for the BCDs, including their two massive members, we use direct metallicity measurements from the literature (Table~\ref{tab:main}). 

\begin{figure}%[ht]
\begin{center}
\includegraphics[height=0.56\textwidth]{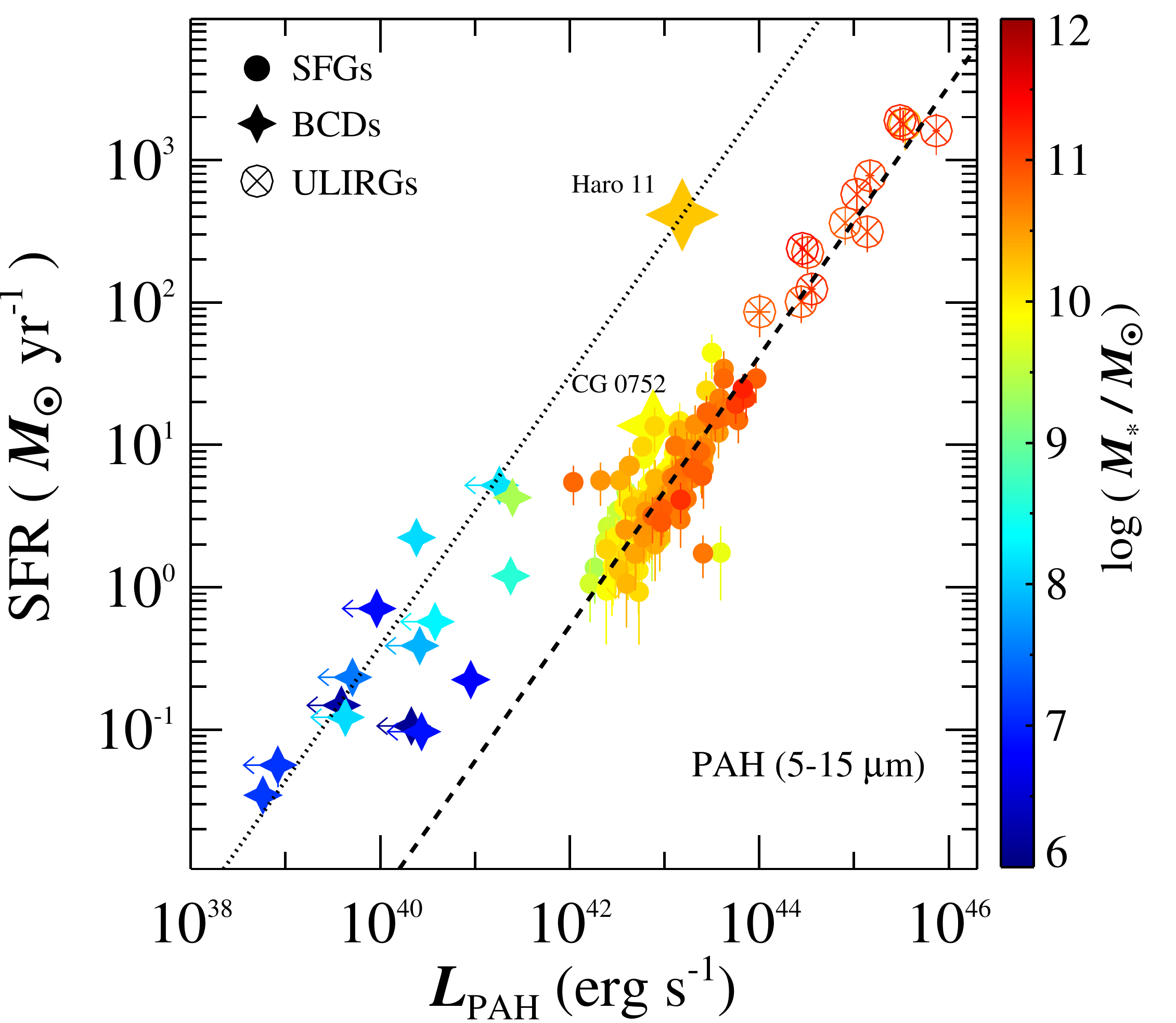} %& 
\caption{The dependence of $L_{\mathrm {PAH}}$ on SFR.  The SFRs are based on $L_{\rm [Ne~{\sc II}]+ [Ne~{\sc III}]}$ (Equation~\ref{eq:ho_sfr}).  The dashed line gives the best-fit regression for the sample of massive ($M_\star \geq 10^9\,M_\odot$) star-forming galaxies and ULIRGs.
The dotted line represents the zero point offset for the low-mass ($M_\star < 10^9\,M_\odot$) BCDs.  Two massive BCDs (Haro~11 and CG~0752) are highlighted. The data points are color-coded according to stellar mass.} 
\label{fig:sfr_pah515}
\end{center}
\end{figure}

Figure~\ref{fig:sfr_pah515} shows the empirical correlation between our neon-based SFRs and the integrated $5-15 \mum$ PAH luminosity.  Massive, star-forming galaxies and ULIRGs trace a tight, linear relation over nearly 4 orders of magnitude in $L_{\mathrm{PAH}}$ and 3 orders of magnitude in SFR.  As anticipated, low-mass BCDs define a distinctly offset sequence with larger scatter.  We parameterize the relation between SFR and PAH luminosity as

\begin{equation}
\log {\rm SFR}\ (M_{\odot}\ {\rm yr}^{-1})\,=\, \alpha \, (\log L_{\rm PAH} - 43) \ + \beta, 
\label{eq:sfr_pah}
\end{equation}

\noindent
where $L_{\rm PAH}$ is in units of erg~s$^{-1}$, $\beta = \beta_h$ for $M_{\ast} \geq 10^9\,M_{\odot}$, and $\beta = \beta_l$ for $M_{\ast} < 10^9\,M_{\odot}$ (Table~\ref{tab:sfr_par}).  We perform regression analysis using the {\tt IDL} procedure {\tt LINMIX\_ERR} (Kelly 2007) to derive the best-fit model parameters and calculate the scatter of the above correlation. Based on Bayesian inference, {\tt LINMIX\_ERR} accounts for measurement errors in the linear regression by computing the likehood function for the observed data, and upper limits in the dependent variable (SFR) are properly treated. The best-fit model parameters are each summarized in the posterior distribution, where a normal density is adopted as priors given the observed dataset. The correlation is almost perfectly linear for massive galaxies (\eg $\alpha = 0.948\pm0.034$ for 5--15 $\mu$m PAH). 

Since $\sim$60\% of the PAH emission in BCDs are upper limits, we constrain the offset in the ${\rm SFR}-L_{\rm PAH}$ relation between BCDs and high-mass galaxies by fixing the slope of the former to that of the latter.  We estimate the offset by bootstrap resampling using two cases, which likely approximate the maximum and minimum offsets of the two galaxy samples. Specifically, we resample 500 times the PAH luminosities of the undetected BCDs, from truncated normal distributions between zero and the 3 $\sigma$ upper limits, or uniform distributions between zero and 3.5 $\sigma$. For the detections, we resample $L_{\rm PAH}$ from normal distributions with $\sigma$ set to the measurement errors. We then multiply the resampled $L_{\rm PAH}$ with $\alpha$ and subtract them from the SFR to estimate the distribution of $\beta_l$; the median and standard deviation of the 500 realizations give the final $\beta_l$  and its 1~$\sigma$ error. The results from the two cases agree within their uncertainties.  The case of the uniform distribution gives $\beta_l - \beta_h = 1.758 \pm 0.083$ for 5--15 $\mu$m PAH. We tabulate the solutions from the uniform distribution in Table 3, and show the average offset on Figure 3.

In addition to the integrated $5-15 \mum$ PAH emission, Table~\ref{tab:sfr_par} also gives the best-fit parameters for several commonly used individual PAH bands at 6.2, 7.7, 8.6, and 11.3$\mum$.  Our PAH-based SFR estimator is applicable to galaxies spanning over 5 orders of magnitude in $M_\star$ ($\sim 10^6 - 10^{11.4}\,M_\odot$) and more than 4 orders of magnitude in SFR ($\sim 0.1 - 2000\,M_\odot\,{\rm yr}^{-1}$), among them dwarf galaxies, main-sequence star-forming galaxies, and vigorous starbursts.

The total scatter of the SFR-$L_{\rm PAH}$ relation for the high-mass sample is quite small, with $\epsilon_h \lesssim 0.2$ dex, unlike low-mass, dwarf galaxies, for which $\epsilon_l \approx 0.7$ dex.  We do not attempt to evaluate the intrinsic scatter because the total scatter is likely dominated by several sources of systematic uncertainty, all of which are difficult to quantify rigorously.  These include uncertainties arising from direct and indirect estimates of the metallicity, assumptions adopted for the calculations of the fractional abundances for neon, extinction correction, which remains non-negligible even in the mid-IR for some heavily obscured galaxies, and unavoidable residual ambiguities from aperture correction when combining IRS spectra taken with different slits (e.g., O'Dowd et al. 2009).  Yet, these complications notwithstanding, the SFR-$L_{\rm PAH}$ relation is remarkably tight for high-mass galaxies.  
Six out of the 17 nearby BCDs suffer from a slight aperture mismatch between the short-low and short-high spectra, but removing these six objects from the sample hardly perturbs the final fits.  Lastly, we verify that the final calibration is also robust with respect to our treatment of upper limits.  Restricting the calibration only to massive galaxies with both neon lines detected yields a fit very similar to that of the full sample that includes upper limits.

We end this section with a cautionary note.  The SFR calibration given above is based on spatially integrated, globally averaged spectra and implicitly assumes that all of the PAH emission is associated exclusively with star-forming regions.  Spatially resolved studies of nearby galaxies reveal an imperfect correspondence between PAH-emitting regions and H~II regions (Maragkoudakis \etal 2018).   PAH emission also arises from inter-arm regions with no obvious association with young stars (see, e.g., Figure~8 in Dale \etal 2009; Crocker \etal 2013), as well as in environments where evolved stellar populations dominate the heating (\eg Bressan \etal 2006; Kaneda \etal 2008).

\begin{table}
\begin{center}
\caption{Best-fit Parameters for the SFR Calibration in PAH Bands}
\label{tab:sfr_par}
%\scriptsize
%\def\arraystretch{0.3}
\begin{tabular}{l r@{$\pm$}l r@{$\pm$}l r@{$\pm$}l r@{$\pm$}l r@{$\pm$}l}% r@{$\pm$}l}
\hline
\hline
\multicolumn{1}{c}{Band} & 
\multicolumn{2}{c}{$\alpha$} & 
\multicolumn{2}{c}{$\beta_{h}$} & 
\multicolumn{2}{c}{$\epsilon_{h}$} &
\multicolumn{2}{c}{$\beta_{l}$} &
\multicolumn{2}{c}{$\epsilon_{l}$} \\  
%\multicolumn{2}{c}{$\xi$}        & 
%\multicolumn{2}{c}{$\epsilon^{l,0}$} \\ 
%\multicolumn{2}{c}{$F_{W4}$} \\
\multicolumn{1}{c}{(1)} & 
\multicolumn{2}{c}{(2)} & 
\multicolumn{2}{c}{(3)} & 
\multicolumn{2}{c}{(4)} & 
\multicolumn{2}{c}{(5)} & 
\multicolumn{2}{c}{(6)} \\ \hline 
%\multicolumn{2}{c}{(7)} & 
%\multicolumn{2}{c}{(8)}\\ \hline 
%### pah band, slope, err, intercept, err,  total scatter sfg, err, total scatter bcd, err, intercept bcd, err ###
PAH~5$-$15$\mum$  &   0.948  &   0.034  &   0.675  &   0.024  &   0.199  &   0.004  &   2.433  &   0.079  &   0.671  &   0.088 \\
PAH~6.2$\mum$     &   1.052  &   0.041  &   2.066  &   0.054  &   0.195  &   0.004  &   3.996  &   0.129  &   0.769  &   0.097 \\
PAH~7.7$\mum$     &   0.999  &   0.036  &   1.134  &   0.026  &   0.161  &   0.004  &    3.150  &   0.128  &   0.739  &   0.095  \\
PAH~8.6$\mum$     &   0.992  &   0.037  &   1.786  &   0.043  &   0.155  &  0.005  &   3.547  &   0.111   &   0.705  &   0.104  \\
PAH~11.3$\mum$    &   1.013  &   0.037  &   1.660  &   0.038  &   0.153  &   0.004  &   3.596  &   0.105   &   0.797  &   0.087 \\
\hline
\hline
\end{tabular}
\end{center}
\tablecomments{%\scriptsize
Best-fit parameters for 
$\log {\rm SFR}\,=\,\alpha\, (\log L_{\rm PAH} - 43)\,+\, \beta$, where $\beta = \beta_h$ for $M_{\ast} \geq 10^9\,M_{\odot}$ and $\beta = \beta_l$ for $M_{\ast} < 10^9\,M_{\odot}$.
Col. (1) PAH band used in calibration.
Col. (2) Slope of the massive galaxies ($M_{\ast} \geq 10^9\,M_{\odot}$).
Col. (3) Zero point of the massive galaxies.
Col. (4) Total scatter of the massive galaxies.
Col. (5) Zero point of the low-mass ($M_{\ast} < 10^9\,M_{\odot}$) dwarf galaxies. 
Col. (6) Total scatter of the dwarf galaxies.
}
\end{table}
%%%%%%%%%%%%%%%%%%%%%%%%%%%%%%%%%%%%%%%%%%%%%%%%%%%%%%%

\subsection{SFR Based on Photometric Bands  \label{subsec:photsfr}}

\begin{figure}%[ht]
\begin{center}
\includegraphics[height=0.45\textheight]{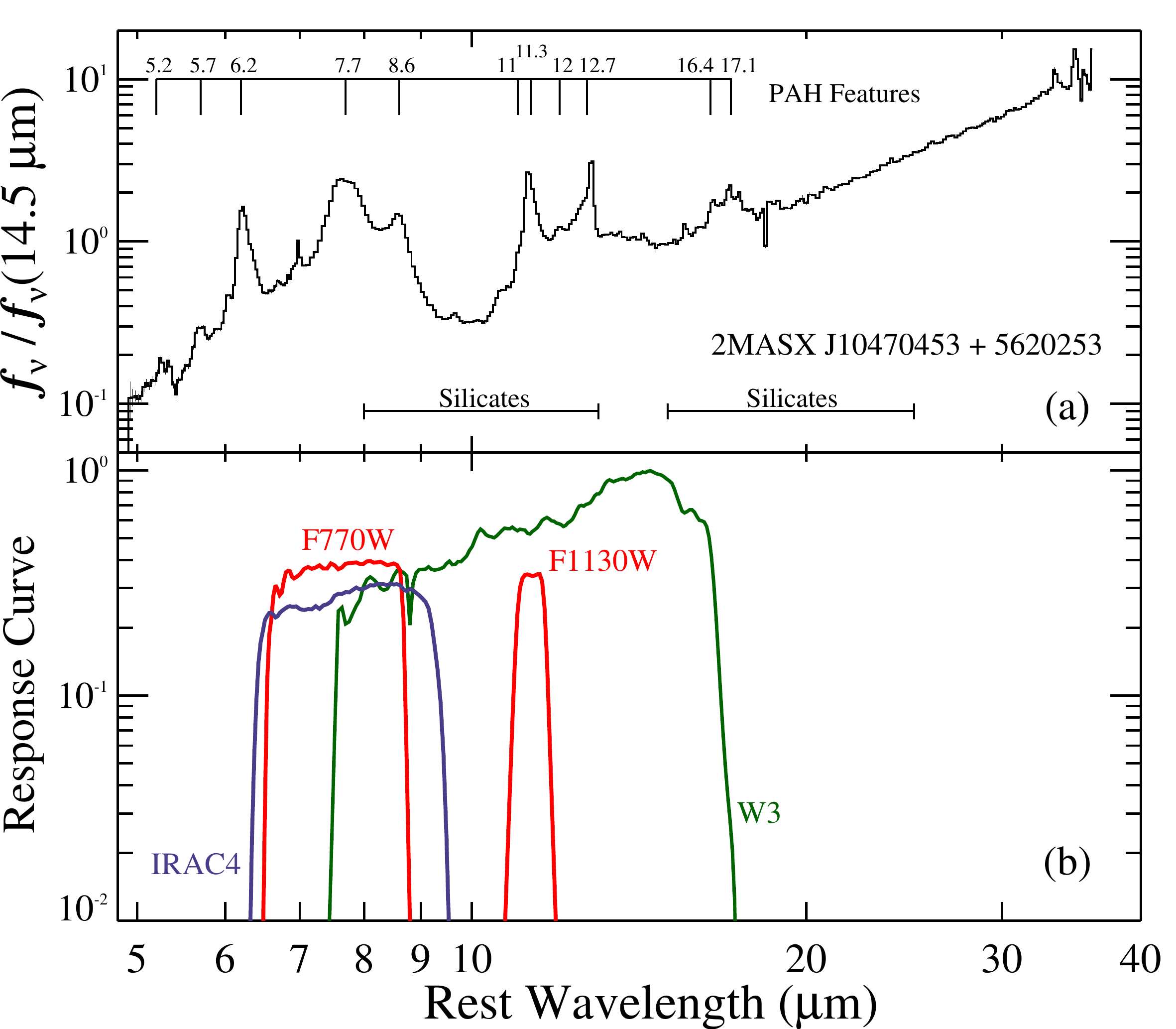} 
\caption{(a) Example mid-IR spectrum of 2MASX~J10470453+5620253, illustrating the locations of the prominent PAH and silicate features.  (b) Response curves of filters that are sensitive to the PAH features, including (green) {\it WISE}\ W3 (12$\mum$) (blue) {\it Spitzer}\ IRAC4 (8$\mum$), and (red) {\it JWST}\ F770W (7.7$\mum$) and F1130W (11.3$\mum$).}
\label{fig:pah_filter}
\end{center}
\end{figure}

PAH emission is sufficiently prominent and widespread throughout the mid-IR spectrum that for local galaxies the more conspicuous features can be captured using the photometric bands of existing and upcoming facilities.  Figure~\ref{fig:pah_filter} highlights some examples applicable to $z\approx 0$ ({\it Spitzer}\ IRAC4 8 $\mum$, {\it WISE}\ W3 12 $\mum$, and {\it JWST}\ F770W at 7.7 $\mum$ and F1130W at 11.3 $\mum$), but obviously appropriately matched bandpasses can be considered for higher redshifts.

%%% Figure 6 PAH_BANDPASS %%%%%%%%%%%%%%%%%%%%%%%%%%%%%%%%%%%%
%%% Figure 6 PAH_BANDPASS %%%%%%%%%%%%%%%%%%%%%%%%%%%%%%%%%%%%%

We calibrate the PAH-sensitive photometric bands against SFRs derived from our extinction-corrected $L_{\rm [Ne~{\sc II}]+ [Ne~{\sc III}]}$, using synthetic photometry obtained by convolving the response curves of the respective filters with the rest-frame spectra of our objects.  For illustration, Figure 5 shows the case of {\it WISE}/W3; the other filters behave qualitatively similarly.  The relation between SFR and the photometric bands is highly nonlinear, even for the massive galaxies alone; a linear fit yields a reduced $\chi^2_\nu = 1.32$ for massive galaxies.  A third-order polynomial gives an improved fit with reduced $\chi^2_\nu = 1.16$.  Instead of treating the massive and dwarf galaxies separately, as we did for the ${\rm SFR}-L_{\rm PAH}$ relation (Figure 3), for simplicity we fit the entire sample with a third-order polynomial,
\begin{equation}
\log {\rm SFR}\ (M_{\odot}\ {\rm yr}^{-1})\,=\,a\,+\, bx\,+\, cx^2\,+\,dx^3 , 
\label{eq:sfr_phot}
\end{equation}
\noindent
where $x\,=\,{\rm \log}\ L_{\rm band} - 43$ and $L_{\rm band}$ is in units of erg~s$^{-1}$. The total scatter is $\sim 0.25-0.3$ dex (Table~\ref{tab:sfr_phot}). 

%
%%%  Figure photometry_PAH  W3 %%%%%%%%%%%%%%%%%%%%%%%%%%%%%%%%
\begin{figure}%[ht]
\begin{center}
\includegraphics[height=0.56\textwidth]{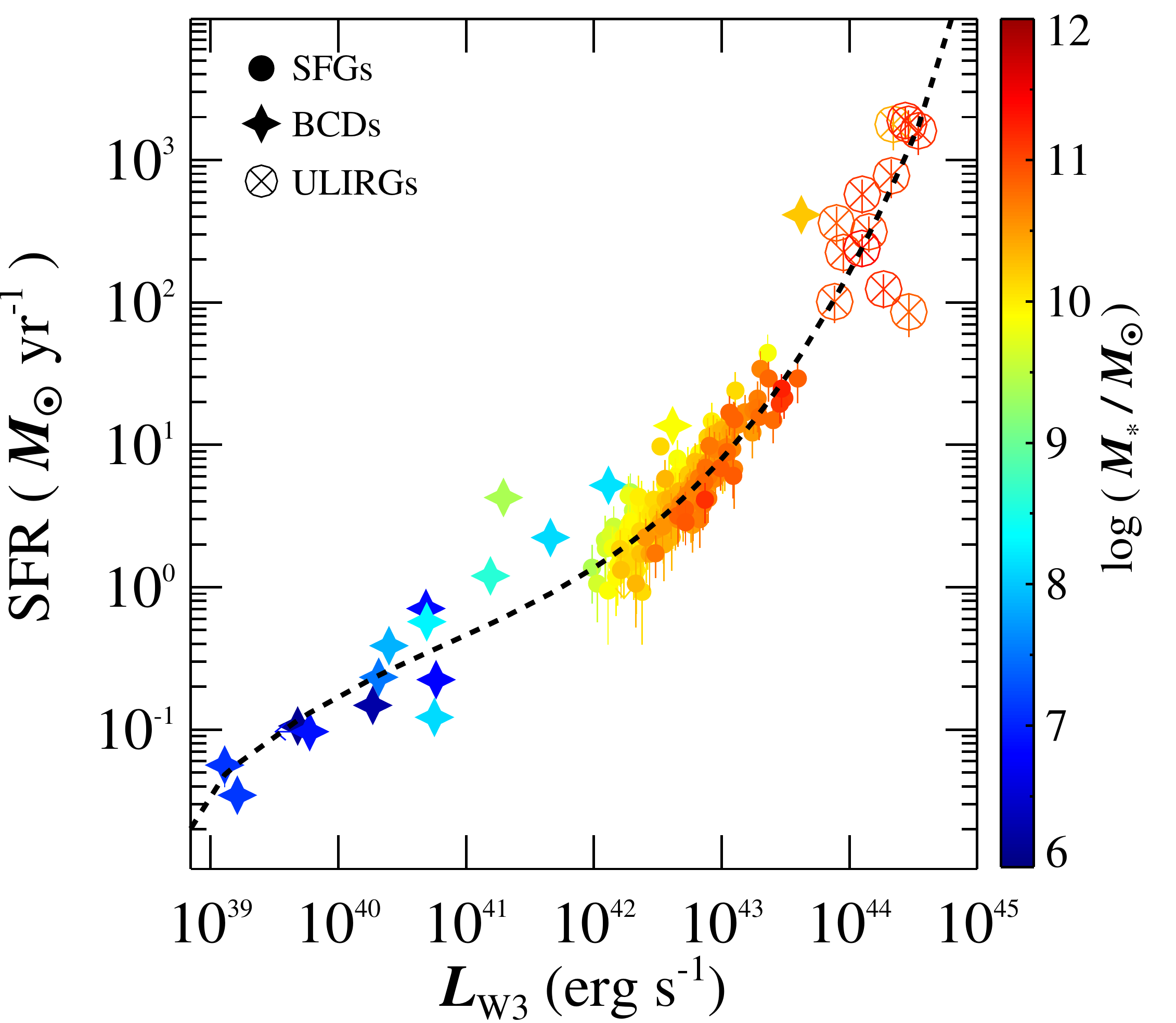} 
\caption {The dependence of $L_{\rm W3}$ on SFR.  The SFRs are based on $L_{\rm [Ne~{\sc II}]+ [Ne~{\sc III}]}$ (Equation~\ref{eq:ho_sfr}). The dashed line gives the best-fit third-order polynomial. The data points are color-coded according to stellar mass.} 
\label{fig:sfr_w3}
\end{center}
\end{figure}

\begin{table}
\begin{center}
\caption{Best-fit Parameters for the SFR Calibration in Photometry Bands}
\label{tab:sfr_phot}
%\scriptsize
%\def\arraystretch{0.3}
\begin{tabular}{l r@{$\pm$}l r@{$\pm$}l r@{$\pm$}l r@{$\pm$}l r@{$\pm$}l}% r@{$\pm$}l}
\hline
\hline
\multicolumn{1}{c}{Band} & 
\multicolumn{2}{c}{$a$} & 
\multicolumn{2}{c}{$b$} & 
\multicolumn{2}{c}{$c$} &
\multicolumn{2}{c}{$d$} &
\multicolumn{2}{c}{$\epsilon$} \\  
%\multicolumn{2}{c}{$\xi$}        & 
%\multicolumn{2}{c}{$\epsilon^{l,0}$} \\ 
%\multicolumn{2}{c}{$F_{W4}$} \\
\multicolumn{1}{c}{(1)} & 
\multicolumn{2}{c}{(2)} & 
\multicolumn{2}{c}{(3)} & 
\multicolumn{2}{c}{(4)} & 
\multicolumn{2}{c}{(5)} & 
\multicolumn{2}{c}{(6)} \\ \hline 
%\multicolumn{2}{c}{(7)} & 
%\multicolumn{2}{c}{(8)}\\ \hline 
%### object, ax3, err_a, bx2, err_b, cx, err_c, d, err_d, tot_scatter, err_tot_scatter ###
IRAC4        &  1.494  &   0.019  &    1.440  &   0.032  &  0.409  &   0.023  &  0.046  &   0.005   &   0.270  &   0.013 \\
F770W        &  1.401  &   0.019  &    1.376  &   0.030  &  0.399  &   0.021  &  0.047  &   0.004   &   0.268  &   0.014 \\
F1130W       &  3.026  &   0.058  &    2.545  &   0.096  &  0.677  &   0.048  &  0.065  &   0.007   &   0.291  &   0.014 \\
W3           &  0.902  &   0.016  &    0.997  &   0.022  &  0.273  &   0.015  &  0.042  &   0.005   &   0.250  &   0.013 \\
\hline
\hline
\end{tabular}
\end{center}
\tablecomments{%\scriptsize
Best-fit parameters for 
                                      $\log {\rm SFR}\,=\,a\,+\, bx\,+\, cx^2\,+\,dx^3$ , where 
                                      $x\,=\,{\rm \log}\ L_{\rm band} - 43$.
Col. (1) Band used in calibration.
Col. (2) Coefficient $a$.
Col. (3) Coefficient $b$.
Col. (4) Coefficient $c$.
Col. (5) Coefficient $d$.
Col. (6) Total scatter.
}
\end{table}
%

%\clearpage 

\section{Discussion\label{sec:discussion}}

\subsection{The PAH Deficit in Dwarf Galaxies\label{subsec:dwarf}}

Most of the BCDs in our sample have conspicuously weak PAH emission, and the degree of PAH deficit is strongly tied to stellar mass. The PAH deficit in Figure 2c is illustrated for the integrated PAH emission from 5--15$\mum$, but we confirm that the effect is qualitatively similar for the PAH 7.7 and 11.3$\mum$ features (see Table~\ref{tab:sfr_par} for quantitative estimates). A number of previous studies have noted the general tendency for dwarf galaxies to exhibit low PAH equivalent widths (\eg Peeters \etal 2004; Engelbracht \etal 2005; Madden \etal 2006; Wu \etal 2006; Shipley \etal 2016).  While the true cause of the reduction in PAH emission is still unknown, the association of weak PAH emission with prominent [Ne~III]/[Ne~II] in low-mass, low-metallicity dwarf galaxies suggests that the harder radiation fields of subsolar metallicity environments may not be conducive to the survival of PAH molecules (\eg Plante \& Sauvage 2002).  Destruction of PAHs by shocks has also been implicated, in view of the elevated levels of [Fe~II] 26$\mum$ emission reported in some systems (O'Halloran \etal 2006).  Perhaps supernova-driven shocks are more prevalent in dwarf galaxies if their low metallicities lead to a more top-heavy initial mass function (\eg Bromm \etal 2002).  However, a careful search for [Fe~II] 26$\mum$ in our sample of BCDs detected this feature in only five out of 17 objects, at a median value of [Fe~II]/[Ne~II] = 0.13 that lies significantly below that of previous observations (O'Halloran \etal 2006). The poor statistics of our sample preclude us from discussing this issue further. %\textbf{Draine \etal 2007; Galliano \etal 2008; Sandstrom \etal 2010; Paradis \etal 2011; Remy-Ruyer \etal 2015}; 

Might the depression of PAH emission be attributed simply to the overall low metal content in dwarf galaxies?  Not likely, not unless carbon is preferentially depleted relative to neon. O'Halloran \etal (2006) speculate that dwarf galaxies may be too young to host a significant population of asymptotic giant branch stars, whose stellar winds are thought to be a main site of dust formation (Morgan \& Edmunds 2003). %, \textbf{despite this process is still in debate (\eg Draine 1990, 2003)}
The ultra-metal-poor dwarf galaxy I~Zw~18 presents a strong counterargument to this hypothesis: its mid-IR spectrum is completely devoid of PAH emission, and yet it is clearly evolved enough to host abundant asymptotic giant branch stars (Izotov \& Thuan 2004).  

We offer an alternative, though related, proposal.  We suggest that PAHs are depleted in dwarf galaxies principally because low-metallicity environments lack sufficient dust grains to shield these large, fragile molecules from photodissociation by UV radiation.  This is analogous to the mechanism commonly invoked to explain the deficit of CO emission in this class of galaxies (\eg Leroy \etal 2011; Magdis \etal 2012; Bolatto \etal 2013).  Quantitative calculations are needed to assess the viability of this hypothesis.

\subsection{Comparison with Previous PAH-Based SFR Calibrations \label{subsec:compare}}

A number of studies have presented empirical calibrations of SFR based on individual PAH bands at 6.2, 7.7, and 11.3$\mum$, or some combinations thereof.  For a proper comparison, we only consider previous works that give calibrations involving the integrated luminosity of the PAH bands, not those based on measurements of peak line intensity (\eg Houck \etal 2007; Weedman \& Houck 2008; Sargsyan \& Weedman 2009), which would be significantly affected by the underlying continuum and extinction.  Furthermore, we convert all calibrations to the Salpeter initial mass function and infer that they pertain to star formation within 20\,Myr, which contributes most of the ionizing flux (Kennicutt 1998).  
Still, quantitative comparison between our work and those of others is fraught with difficulty because of fundamental differences in the manner in which the PAH emission was measured.  For example, whereas our approach (Xie et al. 2018) utilizes global fitting to simultaneously decompose the PAH emission from the underlying continuum components, the PAH measurements of Farrah et al. (2007) and Pope et al. (2008) are based on local continuum fits, and the two approaches can produce quite large differences.  Moreover, our analysis properly accounts for the effects of dust extinction, which can be significant in some systems (e.g., ULIRGs), while these previous studies did not.  Thus, it is virtually impossible to ascertain the significance of the factor of a few difference between our results and those of these studies.

The work of Shipley et al. (2016) offers a better point of comparison.  Shipley et al. used an aperture-matched sample of star-forming galaxies to calibrate the PAH bands at 6.2, 7.7, and 11.3$\mum$, as well as all possible combinations of these three features, against SFRs based on $\rm H{\alpha} + 24\mum$.  Importantly, their analysis measured the PAH strength using the code {\tt PAHFIT} (Smith et al. 2007), which for individual bands compares favorably to the global-fitting method of Xie et al. (2018).  Our SFR calibrations broadly agree with those of Shipley \etal (2016), both in slope and scatter, although our derived SFRs are a factor of 4, 2, and 1.3 higher for PAH 6.2, 7.7, and 11.3$\mum$, respectively. The differences remain even after removing the ULIRGs from the fit. Considering that our calibration sample is more than twice as large as that of Shipley et al., and we adopt different methods for dust extinction correction, we consider this level of discrepancy acceptable.

Cluver \etal (2017) calibrated the \textit{WISE}\ W3 band to SFRs determined from total IR luminosity.  Our W3-based SFRs agree with those of Cluver \etal to within 0.15 dex for normal star-forming galaxies and to $\sim 0.5$ dex for ULIRGs\footnote{The sample of Cluver \etal (2017) has too few dwarf galaxies to permit a meaningful comparison.}. Cluver et al. selected their ULIRGs based on {\it WISE}\ mid-IR colors, which may still contain some degree of contamination by active galactic nuclei (see their Section~3.1).  They also applied a procedure to correct W3 for stellar emission.  Both factors may contribute to the differences with our results. Brown \etal (2017) published a similar calibration for \textit{WISE}/W3 and {\textit Spitzer}/IRAC4 using SFRs tied to extinction-corrected $\rm H{\alpha}$ of 66 local star-forming galaxies. For massive star-forming galaxies, their calibration generally agrees with ours within a factor of 2 for W3 and a factor of 4 for IRAC4, but their prescription for dwarf galaxies is $\sim$ 0.7 dex lower in W3 and 1 dex lower in IRAC4.  The significant discrepancy for dwarf galaxies might be related to the leakage of ionizing photons in low-metallicity environments (\eg Hunter et al. 2010).

\section{Summary\label{sec:summary}}

The prominent, pervasive emission features of PAH have the potential to serve as an effective tracer of star formation in galaxies.  We apply a recently developed mid-IR spectral decomposition technique to quantify the strength of the PAH features using low-resolution {\it Spitzer}/IRS spectra of a diverse sample of 226 star-forming galaxies, ranging from low-mass dwarfs to typical star-forming galaxies and extreme starbursts in ultraluminous IR galaxies.  Together with high-resolution spectra that yield measurements of the [Ne~II] 12.8 $\mu$m and [Ne~III] 15.6 $\mu$m lines, which effectively trace the Lyman continuum of massive stars, we provide a new PAH-based SFR estimator for galaxies.  We present calibrations for the integrated 5--15 $\mu$m PAH emission, the individual features at 6.2, 7.7, 8.6, and 11.3 $\mu$m, and, for completeness, several mid-infrared bandpasses sensitive to PAH.

Our principal results are as follows:

\begin{enumerate}
\item{The extinction-corrected PAH luminosity correlates tightly ($\sim 0.2$ dex) with the neon-based SFR for star-forming galaxies spanning a wide range in stellar masses ($M_{\ast} \approx 10^9 - 10^{11.4}\,M_{\odot}$) and star formation activity ($\mathrm{SFR} \approx 1 - 2000\,M_{\odot}\rm\,yr^{-1}$).}

\item{Low-mass ($M_{\ast} \lesssim 10^{9}\,M_{\odot}$) dwarf galaxies exhibit notably weaker PAH emission relative to the neon lines, as well as larger scatter in their mutual correlation.  Our new SFR estimator can be applied to galaxies in this low-mass branch, down to ${\rm SFR} \approx 0.1\,M_{\odot}\rm\,yr^{-1}$.  We speculate that the PAH deficit in dwarf galaxies originates from photodissociation of PAH molecules in metal-poor, dust-poor environments.}

\item{Our SFRs are broadly consistent with previous PAH SFRs calibrated against total IR luminosity, extinction-corrected $\rm H{\alpha}$, or $\rm H\alpha + 24\mum$ luminosity, but our calibration sample is extended to include more massive, more luminous starbursts.  We also calibrate PAH-sensitive photometric bands against SFRs, for current (\textit{Spitzer}, \textit{WISE}) and future (\textit{JWST}) mid-IR facilities.}
\end{enumerate}

\acknowledgments
We thank an anonymous referee for helpful comments and suggestions. L.C.H. was supported by the National Science Foundation of China (11721303) and the National Key Program for Science and Technology Research and Development (2016YFA0400702).  Y.X. is supported by the China Postdoctoral Science Foundation Grant (2016M591007) and the National Natural Science Foundation of China for Youth Scientist Project (11803001).  Y.X. thanks Li Shao for help in compiling stellar masses from the MPA-JHU catalog; Jinyi Shangguan for providing stellar masses of the BCDs and ULIRGs; Ruancun Li for help with 2MASS photometry for some of the objects; Hassen Yesuf for assistance with statistical analysis; and Sandra Faber, D. Farrah, Robert Kennicutt, Ming-Yang Zhuang, and Lulu Zhang for insightful discussions.  This publication used data products from the Two Micron All-Sky Survey, which is a joint project of the University of Massachusetts and the Infrared Processing and Analysis Center/California Institute of Technology, funded by the National Aeronautics and Space Administration and the National Science Foundation. The Cornell Atlas of \spitzerirs Sources (CASSIS) is a product of the Infrared Science Center at Cornell University, supported by NASA and JPL.
%
%%% Acknowledgements%%%%%%%%%%%%%%%%%%%%%%%%%%%%%%%%%%%%%%%% 
%\clearpage

\appendix

\section{Stellar Masses of ULIRGs and BCDs \label{appendix:mstar_bcd_ulirg}}

%\subsection{ULIRGs \label{appendix:mstar_bcd}}

The stellar masses of ULIRGs are derived as
%%%%%%%%% ULIRG stellar mass %%%%%%%%%%%%%%%%%%%%%
\begin{equation}
\mathrm{log} (\,M_{\ast}/M_\odot\,) = -0.4\,(\,M_J - M_{J, \odot}\,) - 0.75 + 0.34\,(\,B - I\,),
\label{eq:mstar_ulirg}
\end{equation}
\noindent
where $M_J$ and $M_{J, \odot}=3.65$ mag (Blanton \& Roweis 2007) are, respectively, the rest-frame $J$-band absolute magnitudes of the galaxy and the Sun. We take $B - I = 2.0$ mag (Arribas \etal 2004; U \etal 2012; see Shangguan \etal 2019 for more details). The ULIRGs have stellar masses $10^{10.4}-10^{11.4}\,M_{\odot}$.  
%%%%%%%%% ULIRG stellar mass %%%%%%%%%%%%%%%%%%%%%
%
%\subsection{ BCDs \label{appendix:mstar_ulirg}}
%%%%%%%%% BCD stellar mass %%%%%%%%%%%%%%%%%% 
%
For BCDs, the stellar masses are calculated as 

\begin{equation}
\mathrm{log} (\,M_{\ast}/M_\odot\,) =  -0.4\,(\,M_J - M_{J, \odot}\,) - 0.153 + 0.283\,\langle g-i \rangle\,,
\label{eq:mstar_bcd}
\end{equation} 

\noindent
where $\langle g-i \rangle$\,=\,0.738 mag, the color of the low-surface brightness component as given in Meyer \etal (2014).  The typical mass uncertainty derived in this manner is 0.2 dex (Conroy 2013; Zhang \etal 2016). 
 
Four BCDs are not archived in either the 2MASS extended-source catalog or the point-source catalog because of their low surface brightness\footnote{%
See https://old.ipac.caltech.edu/2mass/releases/allsky/ for more details.}. 
We measure the $J-$band photometry of these sources using our own procedure for galaxy photometry (R. Li \etal, in preparation).  All of them have $J-$band signal-noise-ratio $< 5$, below the criterion to be included in published catalogs. 
SBS~1415+437 is detected in the $J$ band at the 2.5$\sigma$ level; we can only obtain an upper limit for PAH, but both [Ne II] and [Ne III] are significant on detected.
%%%%%%%%% BCD stellar mass %%%%%%%%%%%%%%%%%%%%%

%%%%%%%%%%%%%%%%%%%%%%%%%%%%%%%%%%%%%%%%%%%%%%%%%%%%%%%%%


\clearpage

\begin{thebibliography}{}
\expandafter\ifx\csname natexlab\endcsname\relax\def\natexlab#1{#1}\fi

\bibitem[Alonso-Herrero et al.(2006)]{2006ApJ...650..835A} Alonso-Herrero, A., Rieke, G.~H., Rieke, M.~J., et al.\ 2006, \apj, 650, 835

\bibitem[Arribas et al.(2004)]{Arribas2004AJ} Arribas, S., Bushouse, H., Lucas, R.~A., Colina, L., \& Borne, K.~D.\ 2004, \aj, 127, 2522

\bibitem[Armus et al.(2007)]{2007ApJ...656..148A} Armus, L., Charmandaris, V., Bernard-Salas, J., et al.\ 2007, \apj, 656, 148 

\bibitem[Bavouzet et al.(2008)]{2008A&A...479...83B} Bavouzet, N., Dole, H., Le Floc'h, E., et al.\ 2008, \aap, 479, 83 

\bibitem[Bell et al.(2003)]{2003ApJS..149..289B} Bell, E.~F., McIntosh, D.~H., Katz, N., \& Weinberg, M.~D.\ 2003, \apjs, 149, 289

\bibitem[Bergvall et al.(2000)]{2000A&A...359...41B} Bergvall, N., Masegosa, J., {\"O}stlin, G., \& Cernicharo, J.\ 2000, \aap, 359, 41 

\bibitem[Blanton \& Roweis(2007)]{Blanton2007AJ} Blanton, M.~R., \& Roweis, S. 2007, \aj, 133, 734

\bibitem[Bolatto et al.(2013)]{2013ARA&A..51..207B} Bolatto, A.~D., Wolfire, M., \& Leroy, A.~K.\ 2013, \araa, 51, 207 

\bibitem[Brada{\v{c}} et al.(2017)]{2017ApJ...836L...2B} Brada{\v{c}}, M., Garcia-Appadoo, D., Huang, K.-H., et al.\ 2017, \apjl, 836, L2	 
\bibitem[Bressan et al.(2006)]{2006ApJ...639L..55B} Bressan, A., Panuzzo, P., Buson, L., et al.\ 2006, \apjl, 639, L55 

\bibitem[Bromm et al.(2002)]{2002ApJ...564...23B} Bromm, V., Coppi, P.~S., \& Larson, R.~B.\ 2002, \apj, 564, 23

\bibitem[Brown et al.(2017)]{2017ApJ...847..136B} Brown, M.~J.~I., Moustakas, J., Kennicutt, R.~C., et al.\ 2017, \apj, 847, 136 

\bibitem[Calzetti(2001)]{2001PASP..113.1449C} Calzetti, D.\ 2001, \pasp, 113, 1449
 
\bibitem[Calzetti(2013)]{2013seg..book..419C} Calzetti, D.\ 2013, Secular Evolution of Galaxies, 419
 
\bibitem[Calzetti et al.(2010)]{2010ApJ...714.1256C} Calzetti, D., Wu, S.-Y., Hong, S., et al.\ 2010, \apj, 714, 1256
 
\bibitem[Cluver et al.(2017)]{2017ApJ...850...68C} Cluver, M.~E., Jarrett, T.~H., Dale, D.~A., et al.\ 2017, \apj, 850, 68 

\bibitem[Conroy(2013)]{Conroy2013ARAA} Conroy, C.\ 2013, \araa, 51, 393 
 
\bibitem[Crocker et al.(2013)]{2013ApJ...762...79C} Crocker, A.~F., Calzetti, D., Thilker, D.~A., et al.\ 2013, \apj, 762, 79

\bibitem[Dale et al.(2009)]{2009ApJ...703..517D} Dale, D.~A., Cohen, S.~A., Johnson, L.~C., et al.\ 2009, \apj, 703, 517 

\bibitem[De Looze et al.(2011)]{2011MNRAS.416.2712D} De Looze, I., Baes, M., Bendo, G.~J., Cortese, L., \& Fritz, J.\ 2011, \mnras, 416, 2712

\bibitem[Diamond-Stanic \& Rieke(2012)]{2012ApJ...746..168D} Diamond-Stanic, A.~M., \& Rieke, G.~H.\ 2012, \apj, 746, 168 

\bibitem[Draine \& Li(2001)]{2001ApJ...551..807D} Draine, B.~T., \& Li, A.\ 2001, \apj, 551, 807 

\bibitem[Draine \& Li(2007)]{2007ApJ...657..810D} Draine, B.~T., \& Li, A.\ 2007, \apj, 657, 810 

\bibitem[Engelbracht et al.(2005)]{2005ApJ...628L..29E} Engelbracht, C.~W., Gordon, K.~D., Rieke, G.~H., et al.\ 2005, \apjl, 628, L29

\bibitem[Farrah et al.(2007)]{2007ApJ...667..149F} Farrah, D., Bernard-Salas, J., Spoon, H.~W.~W., et al.\ 2007, \apj, 667, 149 

\bibitem[Ferland et al.(2017)]{2017RMxAA..53..385F} Ferland, G.~J., Chatzikos, M., Guzm{\'a}n, F., et al.\ 2017, \rmxaa, 53, 385 

\bibitem[Ferland et al.(1998)]{1998PASP..110..761F} Ferland, G.~J., Korista, K.~T., Verner, D.~A., et al.\ 1998, \pasp, 110, 761 

\bibitem[F{\"o}rster Schreiber et al.(2004)]{2004A&A...419..501F} F{\"o}rster Schreiber, N.~M., Roussel, H., Sauvage, M., \& Charmandaris, V.\ 2004, \aap, 419, 501 

\bibitem[Gallagher et al.(1989)]{1989AJ.....97..700G} Gallagher, J.~S., Bushouse, H., \& Hunter, D.~A.\ 1989, \aj, 97, 700

\bibitem[Hao et al.(2011)]{2011ApJ...741..124H} Hao, C.-N., Kennicutt, R.~C., Johnson, B.~D., et al.\ 2011, \apj, 741, 124

\bibitem[Hirashita et al.(2003)]{2003A&A...410...83H} Hirashita, H., Buat, V., \& Inoue, A.~K.\ 2003, \aap, 410, 83

\bibitem[Ho \& Keto(2007)]{2007ApJ...658..314H} Ho, L.~C., \& Keto, E.\ 2007, \apj, 658, 314 

\bibitem[Houck et al.(2004)]{2004SPIE.5487...62H} Houck, J.~R., Roellig, T.~L., Van Cleve, J., et al.\ 2004, \procspie, 5487, 62 

\bibitem[Houck et al.(2007)]{2007ApJ...671..323H} Houck, J.~R., Weedman, D.~W., Le Floc'h, E., \& Hao, L.\ 2007, \apj, 671, 323 

\bibitem[Hunter et al.(2010)]{2010AJ....139..447H} Hunter, D.~A., Elmegreen, B.~G., \& Ludka, B.~C.\ 2010, \aj, 139, 447 

\bibitem[Iwasawa et al.(2011)]{2011A&A...529A.106I} Iwasawa, K., Sanders, D.~B., Teng, S.~H., et al.\ 2011, \aap, 529, A106 

\bibitem[Izotov et al.(1997)]{1997ApJ...476..698I} Izotov, Y.~I., Lipovetsky, V.~A., Chaffee, F.~H., et al.\ 1997, \apj, 476, 698 

\bibitem[Izotov \& Thuan(1999)]{1999ApJ...511..639I} Izotov, Y.~I., \& Thuan, T.~X.\ 1999, \apj, 511, 639 

\bibitem[Izotov \& Thuan(2004)]{2004ApJ...616..768I} Izotov, Y.~I., \& Thuan, T.~X.\ 2004, \apj, 616, 768 

\bibitem[Izotov et al.(2012)]{2012MNRAS.427.1229I} Izotov, Y.~I., Thuan, T.~X., \& Privon, G.\ 2012, \mnras, 427, 1229 

\bibitem[Jimmy et al.(2015)]{2015ApJ...812...98J} Jimmy, Tran, K.-V., Saintonge, A., et al.\ 2015, \apj, 812, 98 

\bibitem[Kaneda et al.(2008)]{2008ApJ...684..270K} Kaneda, H., Onaka, T., Sakon, I., et al.\ 2008, \apj, 684, 270 

\bibitem[Kelly(2007)]{2007ApJ...665.1489K} Kelly, B.~C.\ 2007, \apj, 665, 1489 

\bibitem[Kennicutt(1998)]{1998ARA&A..36..189K} Kennicutt Jr., R.~C. 1998, \araa, 36, 189 

\bibitem[Kennicutt \& Evans(2012)]{2012ARA&A..50..531K} Kennicutt Jr., R.~C., \& Evans, N.~J.\ 2012, \araa, 50, 531 

\bibitem[Kennicutt et al.(2009)]{2009ApJ...703.1672K} Kennicutt Jr., R.~C., Hao, C.-N., Calzetti, D., et al.\ 2009, \apj, 703, 1672 

\bibitem[Kewley \& Dopita(2002)]{2002ApJS..142...35K} Kewley, L.~J., \& Dopita, M.~A.\ 2002, \apjs, 142, 35 

\bibitem[Kewley \& Ellison(2008)]{2008ApJ...681.1183K} Kewley, L.~J., \& Ellison, S.~L.\ 2008, \apj, 681, 1183 

\bibitem[Kobulnicky \& Skillman(1996)]{1996ApJ...471..211K} Kobulnicky, H.~A., \& Skillman, E.~D.\ 1996, \apj, 471, 211 

\bibitem[Kroupa \& Weidner(2003)]{2003ApJ...598.1076K} Kroupa, P., \& Weidner, C.\ 2003, \apj, 598, 1076 

\bibitem[LaMassa et al.(2012)]{2012ApJ...758....1L} LaMassa, S.~M., Heckman, T.~M., Ptak, A., et al.\ 2012, \apj, 758, 1

\bibitem[Lebouteiller et al.(2015)]{2015ApJS..218...21L} Lebouteiller, V., Barry, D.~J., Goes, C., et al.\ 2015, \apjs, 218, 21

\bibitem[Lebouteiller et al.(2011)]{2011ApJS..196....8L} Lebouteiller, V., Barry, D.~J., Spoon, H.~W.~W., et al.\ 2011, \apjs, 196, 8 

\bibitem[Leroy et al.(2011)]{2011ApJ...737...12L} Leroy, A.~K., Bolatto, A., Gordon, K., et al.\ 2011, \apj, 737, 12 

\bibitem[Madau \& Dickinson(2014)]{2014ARA&A..52..415M} Madau, P., \& Dickinson, M.\ 2014, \araa, 52, 415

\bibitem[Madden et al.(2006)]{2006A&A...446..877M} Madden, S.~C., Galliano, F., Jones, A.~P., \& Sauvage, M.\ 2006, \aap, 446, 877 

\bibitem[Magdis et al.(2012)]{2012ApJ...760....6M} Magdis, G.~E., Daddi, E., B{\'e}thermin, M., et al.\ 2012, \apj, 760, 6 

\bibitem[Maragkoudakis et al.(2018)]{2018MNRAS.481.5370M} Maragkoudakis, A., Ivkovich, N., Peeters, E., et al.\ 2018, \mnras, 481, 5370 

\bibitem[Markwardt(2009)]{2009ASPC..411..251M} Markwardt, C.~B.\ 2009, Astronomical Data Analysis Software and Systems XVIII, 411, 251

\bibitem[Mathis et al.(1983)]{1983A&A...128..212M} Mathis, J.~S., Mezger, P.~G., \& Panagia, N.\ 1983, \aap, 128, 212 

\bibitem[Meyer et al.(2014)]{2014A&A...562A..49M} Meyer, H.~T., Lisker, T., Janz, J., \& Papaderos, P.\ 2014, \aap, 562, A49 

\bibitem[Morgan \& Edmunds(2003)]{2003MNRAS.343..427M} Morgan, H.~L., \& Edmunds, M.~G.\ 2003, \mnras, 343, 427

\bibitem[Mould et al.(2000)]{2000ApJ...529..786M} Mould, J.~R., Huchra, J.~P., Freedman, W.~L., et al.\ 2000, \apj, 529, 786

\bibitem[O'Dowd et al.(2009)]{2009ApJ...705..885O} O'Dowd, M.~J., Schiminovich, D., Johnson, B.~D., et al.\ 2009, \apj, 705, 885 

\bibitem[O'Dowd et al.(2011)]{2011ApJ...741...79O} O'Dowd, M.~J., Schiminovich, D., Johnson, B.~D., et al.\ 2011, \apj, 741, 79

\bibitem[O'Halloran et al.(2006)]{2006ApJ...641..795O} O'Halloran, B., Satyapal, S., \& Dudik, R.~P.\ 2006, \apj, 641, 795 

\bibitem[Peeters et al.(2004)]{2004ApJ...613..986P} Peeters, E., Spoon, H.~W.~W., \& Tielens, A.~G.~G.~M.\ 2004, \apj, 613, 986

\bibitem[Pentericci et al.(2016)]{2016ApJ...829L..11P} Pentericci, L., Carniani, S., Castellano, M., et al.\ 2016, \apjl, 829, L11

\bibitem[P{\'e}rez-Gonz{\'a}lez et al.(2006)]{2006ApJ...648..987P} P{\'e}rez-Gonz{\'a}lez, P.~G., Kennicutt, R.~C., Jr., Gordon, K.~D., et al.\ 2006, \apj, 648, 987 

\bibitem[Pettini \& Pagel(2004)]{2004MNRAS.348L..59P} Pettini, M., \& Pagel, B.~E.~J.\ 2004, \mnras, 348, L59

\bibitem[Planck Collaboration et al.(2016)]{2016A&A...594A..13P} Planck Collaboration, Ade, P.~A.~R., Aghanim, N., et al.\ 2016, \aap, 594, A13

\bibitem[Plante \& Sauvage(2002)]{2002AJ....124.1995P} Plante, S., \& Sauvage, M.\ 2002, \aj, 124, 1995 

\bibitem[Pope et al.(2008)]{2008ApJ...675.1171P} Pope, A., Chary, R.-R., Alexander, D.~M., et al.\ 2008, \apj, 675, 1171 


\bibitem[Rela{\~n}o et al.(2012)]{2012MNRAS.423.2933R} Rela{\~n}o, M., Kennicutt, R.~C., Eldridge, J.~J., et al.\ 2012, \mnras, 423, 2933

\bibitem[Rela{\~n}o et al.(2007)]{2007ApJ...667L.141R} Rela{\~n}o, M., Lisenfeld, U., P{\'e}rez-Gonz{\'a}lez, P.~G., V{\'{\i}}lchez, J.~M., \& Battaner, E.\ 2007, \apjl, 667, L141 

\bibitem[Rieke et al.(2009)]{2009ApJ...692..556R} Rieke, G.~H., Alonso-Herrero, A., Weiner, B.~J., et al.\ 2009, \apj, 692, 556 

\bibitem[Salpeter(1955)]{1955ApJ...121..161S} Salpeter, E.~E.\ 1955, \apj, 121, 161 

\bibitem[Sargsyan \& Weedman(2009)]{2009ApJ...701.1398S} Sargsyan, L.~A., \& Weedman, D.~W.\ 2009, \apj, 701, 1398

\bibitem[Searle \& Sargent(1972)]{1972ApJ...173...25S} Searle, L., \& Sargent, W.~L.~W.\ 1972, \apj, 173, 25 

\bibitem[Shangguan et al.(2018)]{} Shangguan, J., Ho, L.~C., \& Li, R., et al.\ 2019, \apj, 870, 104

\bibitem[Shipley et al.(2013)]{2013ApJ...769...75S} Shipley, H.~V., Papovich, C., Rieke, G.~H., et al.\ 2013, \apj, 769, 75 

\bibitem[Shipley et al.(2016)]{2016ApJ...818...60S} Shipley, H.~V., Papovich, C., Rieke, G.~H., Brown, M.~J.~I., \& Moustakas, J.\ 2016, \apj, 818, 60 

\bibitem[Smith et al.(2007)]{2007ApJ...656..770S} Smith, J.~D.~T., Draine, B.~T., Dale, D.~A., et al.\ 2007, \apj, 656, 770 

\bibitem[Stacey et al.(1991)]{1991ApJ...373..423S} Stacey, G.~J., Geis, N., Genzel, R., et al.\ 1991, \apj, 373, 423 

\bibitem[Stacey et al.(2010)]{2010ApJ...724..957S} Stacey, G.~J., Hailey-Dunsheath, S., Ferkinhoff, C., et al.\ 2010, \apj, 724, 957 

\bibitem[Stern et al.(2012)]{2012ApJ...753...30S} Stern, D., Assef, R.~J., Benford, D.~J., et al.\ 2012, \apj, 753, 30

\bibitem[Thornley et al.(2000)]{2000ApJ...539..641T} Thornley, M.~D., Schreiber, N.~M.~F., Lutz, D., et al.\ 2000, \apj, 539, 641 

\bibitem[Torres-Alb{\`a} et al.(2018)]{2018arXiv181002371T} Torres-Alb{\`a}, N., Iwasawa, K., D{\'{\i}}az-Santos, T., et al.\ 2018, A\&A, 620, A140

\bibitem[Tremonti et al.(2004)]{2004ApJ...613..898T} Tremonti, C.~A., Heckman, T.~M., Kauffmann, G., et al.\ 2004, \apj, 613, 898 

\bibitem[Treyer et al.(2010)]{2010ApJ...719.1191T} Treyer, M., Schiminovich, D., Johnson, B.~D., et al.\ 2010, \apj, 719, 1191 

\bibitem[U et al.(2012)]{U2012ApJS} U, V., Sanders, D.~B., Mazzarella, J.~M., et 
al.\ 2012, \apjs, 203, 9 

\bibitem[Vallini et al.(2015)]{2015ApJ...813...36V} Vallini, L., Gallerani, S., Ferrara, A., Pallottini, A., \& Yue, B.\ 2015, \apj, 813, 36

\bibitem[Veilleux et al.(1999)]{1999ApJ...522..113V} Veilleux, S., Kim, D.-C., \& Sanders, D.~B.\ 1999, \apj, 522, 113 

\bibitem[Veilleux et al.(1995)]{1995ApJS...98..171V} Veilleux, S., Kim, D.-C., Sanders, D.~B., Mazzarella, J.~M., \& Soifer, B.~T.\ 1995, \apjs, 98, 171 

\bibitem[Weedman \& Houck(2008)]{2008ApJ...686..127W} Weedman, D.~W., \& Houck, J.~R.\ 2008, \apj, 686, 127 

\bibitem[Werner et al.(2004)]{2004ApJS..154....1W} Werner, M.~W., Roellig, T.~L., Low, F.~J., et al.\ 2004, \apjs, 154, 1 

\bibitem[Wu et al.(2005)]{2005ApJ...632L..79W} Wu, H., Cao, C., Hao, C.-N., et al.\ 2005, \apjl, 632, L79 

\bibitem[Wu et al.(2006)]{2006ApJ...639..157W} Wu, Y., Charmandaris, V., Hao, L., et al.\ 2006, \apj, 639, 157 

\bibitem[Xie et al.(2018)]{2018ApJ...860..154X} Xie, Y., Ho, L.~C., Li, A., \& Shangguan, J.\ 2018, \apj, 860, 154 

\bibitem[Yates et al.(2019)]{2019arXiv190102890Y} Yates, R.~M., Schady, P., Chen, T.-W., Schweyer, T., \& Wiseman, P.\ 2019, \mnras, submitted (arXiv:1901.02890)

\bibitem[Zhang et al.(2016)]{2016ApJ...819L..27Z} Zhang, Z., Shi, Y., Rieke, G.~H., et al.\ 2016, \apjl, 819, L27  

\bibitem[Zhuang et al.(2019)]{2019arXiv190203849Z} Zhuang, M.-Y., Ho, L.~C., \& Shangguan, J.\ 2019, \apj, 873, 103

%%%%%%%%%%%%%%%%%%%%%%%%%%%%%%%%%%%%%%%%%%%%%%%%%%%%%%%%%%%%%%%%%%%%%%%%%%%%%%%%%%%%%%%%%%%%%%%%%%%%%%%%%%%%%%%%%%%%%%%%%%%%%%%%%%%%%%%
%\bibitem[Battisti et al.(2015)]{2015ApJ...800..143B} Battisti, A.~J., Calzetti, D., Johnson, B.~D., \& Elbaz, D.\ 2015, \apj, 800, 143
%\bibitem[Berg et al.(2016)]{2016ApJ...827..126B} Berg, D.~A., Skillman, E.~D., Henry, R.~B.~C., Erb, D.~K., \& Carigi, L.\ 2016, \apj, 827, 126 
%\bibitem[Binder \& Povich(2018)]{2018ApJ...864..136B} Binder, B.~A., \& Povich, M.~S.\ 2018, \apj, 864, 136 
%\bibitem[Boselli et al.(2002)]{2002A&A...385..454B} Boselli, A., Gavazzi, G., Lequeux, J., \& Pierini, D.\ 2002, \aap, 385, 454 
%\bibitem[Brown et al.(2014)]{2014ApJS..212...18B} Brown, M.~J.~I., Moustakas, J., Smith, J.-D.~T., et al.\ 2014, \apjs, 212, 18
%\textcolor{red}{
%\bibitem[Cappellari, \& Copin(2003)]{2003MNRAS.342..345C} Cappellari, M., \& Copin, Y.\ 2003, \mnras, 342, 345}
%\bibitem[Cluver et al.(2014)]{2014ApJ...782...90C} Cluver, M.~E., Jarrett, T.~H., Hopkins, A.~M., et al.\ 2014, \apj, 782, 90 
%\bibitem[Crone et al.(2002)]{2002ApJ...567..258C} Crone, M.~M., Schulte-Ladbeck, R.~E., Greggio, L., \& Hopp, U.\ 2002, \apj, 567, 258
%\bibitem[Davies et al.(2015)]{2015MNRAS.452..616D} Davies, L.~J.~M., Robotham, A.~S.~G., Driver, S.~P., et al.\ 2015, \mnras, 452, 616
%\bibitem[Desai et al.(2007)]{2007ApJ...669..810D} Desai, V., Armus, L., Spoon, H.~W.~W., et al.\ 2007, \apj, 669, 810 
%\bibitem[D{\'{\i}}az-Santos et al.(2013)]{2013ApJ...774...68D} D{\'{\i}}az-Santos, T., Armus, L., Charmandaris, V., et al.\ 2013, \apj, 774, 68 
%\bibitem[Dors et al.(2011)]{2011MNRAS.415.3616D} Dors, O.~L., Jr., Krabbe, A., H{\"a}gele, G.~F., \& P{\'e}rez-Montero, E.\ 2011, \mnras, 415, 3616 
%\bibitem[Draine(1990)]{1990ASPC...12..193D} Draine, B.~T.\ 1990, The Evolution of the Interstellar Medium, 12, 193 
%\bibitem[Draine(2003)]{2003ARA&A..41..241D} Draine, B.~T.\ 2003, \araa, 41, 241
%\bibitem[Wright et al.(2010)]{2010AJ....140.1868W} Wright, E.~L., Eisenhardt, P.~R.~M., Mainzer, A.~K., et al.\ 2010, \aj, 140, 1868 
%\bibitem[Vermeij et al.(2002)]{2002A&A...382.1042V} Vermeij, R., Peeters, E., Tielens, A.~G.~G.~M., \& van der Hulst, J.~M.\ 2002, \aap, 382, 1042 
%\bibitem[Murphy et al.(2011)]{2011ApJ...737...67M} Murphy, E.~J., Condon, J.~J., Schinnerer, E., et al.\ 2011, \apj, 737, 67
%\bibitem[Senarath et al.(2018)]{2018arXiv181200534S} Senarath, M.~R., Brown, M.~J.~I., Cluver, M.~E., et al.\ 2018, \apj, 869, L26
%\bibitem[Sargsyan et al.(2012)]{2012ApJ...755..171S} Sargsyan, L., Lebouteiller, V., Weedman, D., et al.\ 2012, \apj, 755, 171 
%\bibitem[Rodriguez-Fernandez et al.(2006)]{2006A&A...453...77R} Rodriguez-Fernandez, N.~J., Braine, J., Brouillet, N., \& Combes, F.\ 2006, \aap, 453, 77 
%\bibitem[Salim et al.(2018)]{2018ApJ...859...11S} Salim, S., Boquien, M., \& Lee, J.~C.\ 2018, \apj, 859, 11
%\bibitem[Press et al.(1992)]{1992nrca.book.....P} Press, W.~H., Teukolsky, S.~A., Vetterling, W.~T., \& Flannery, B.~P.\ 1992, in Numerical Recipes in C: The Art of Scientific Computing (2nd ed.; Cambridge: Cambridge Univ. Press)

%\bibitem[Heesen et al.(2014)]{2014AJ....147..103H} Heesen, V., Brinks, E., Leroy, A.~K., et al.\ 2014, \aj, 147, 103
%\bibitem[Rubin, D. B. (1992)]{} Gelman, A., \& Rubin, D. B. (1992), Inference from Iterative Simulation Using Multiple Sequences, Statistical Science, 7, 457-511 
%\bibitem[Gelman et al. (2004)]{} Gelman, A., Carlin, J. B., Stern, H. S., \& Rubin, D. B. 2004, Bayesian Data Analysis (2nd ed.; Boca Raton: Chapman \& Hall)
%\bibitem[Galliano(2004)]{2004PhDT.........8G} Galliano, F.\ 2004, Ph.D.~Thesis, Universit\`e de Paris XI
%\bibitem[Hao et al.(2009)]{2009ApJ...704.1159H} Hao, L., Wu, Y., Charmandaris, V., et al.\ 2009, \apj, 704, 1159 
%\bibitem[Ho(2005)]{2005ApJ...629..680H} Ho, L.~C.\ 2005, \apj, 629, 680
%\bibitem[Huang et al.(2009)]{2009ApJ...700..183H} Huang, J.-S., Faber, S.~M., Daddi, E., et al.\ 2009, \apj, 700, 183 
%\bibitem[Kim et al.(2006)]{2006ApJ...642..702K} Kim, M., Ho, L.~C., \& Im, M.\ 2006, \apj, 642, 702 
%\bibitem[Leitherer et al.(1999)]{1999ApJS..123....3L} Leitherer, C., Schaerer, D., Goldader, J.~D., et al.\ 1999, \apjs, 123, 3 
%\bibitem[Malhotra et al.(1997)]{1997ApJ...491L..27M} Malhotra, S., Helou, G., Stacey, G., et al.\ 1997, \apjl, 491, L27 

\end{thebibliography}
\end{document}